\documentclass[12pt, a4paper]{article}

\usepackage{multirow}
\usepackage{amsmath}
\usepackage{amsfonts}
\usepackage{amssymb}
\usepackage{graphicx, rotating}
\usepackage{epstopdf}
\usepackage{epsfig}
\usepackage{latexsym}
\usepackage{graphicx}
\usepackage{color}
\usepackage{amsmath,amssymb}
\usepackage{cite}
\usepackage{slashed}

\ifx\pdfoutput\undefined
\usepackage[dvips,bookmarks=false]{hyperref}	
\else
\usepackage{hyperref}	
\fi

\hypersetup{colorlinks, citecolor=bluscuro, linkcolor=black, urlcolor=bluscuro}
\definecolor{rossos}{cmyk}{0,1,1,0.55}
\definecolor{bluscuro}{rgb}{0.15, 0.2, .85}
\definecolor{bluchiaro}{cmyk}{1,.3,0.,0.1}

\newcommand{\fref}[1]{Fig.~\ref{fig:#1}} 
\newcommand{\eref}[1]{Eq.~\eqref{eq:#1}}

\newcommand{\sref}[1]{Section~\ref{sec:#1}}
\newcommand{\cref}[1]{Chapter~\ref{ch:.#1}}
\newcommand{\tref}[1]{Table~\ref{tab:#1}}

\newcommand{\beq}{\begin{equation}} 
\newcommand{\eeq}{\end{equation}} 
\newcommand{\ba}{\begin{array}}  
\newcommand{\ea}{\end{array}} 
\newcommand{\bea}{\begin{eqnarray}}  
\newcommand{\eea}{\end{eqnarray} }  
\newcommand{\be}{\begin{eqnarray}}  
\newcommand{\ee}{\end{eqnarray} }  
\newcommand{\bal}{\begin{align}}
\newcommand{\eal}{\end{align}}   
\newcommand{\bi}{\begin{itemize}}  
\newcommand{\ei}{\end{itemize}}  
\newcommand{\ben}{\begin{enumerate}}  
\newcommand{\een}{\end{enumerate}}  
\newcommand{\bc}{\begin{center}}
\newcommand{\ec}{\end{center}} 
\newcommand{\bt}{\begin{table}}
\newcommand{\et}{\end{table}}  
\newcommand{\btb}{\begin{tabular}}
\newcommand{\etb}{\end{tabular}}  
\newcommand{\bvec}{\left ( \ba{c}}
\newcommand{\evec}{\ea \right )}

\newcommand{\cO}{{\mathcal O}} 
 
\newcommand{\cL}{{\mathcal L}}

\newcommand{\gev}{\mathrm{GeV}}
\newcommand{\tev}{\mathrm{TeV}}

\def\hc{{\rm h.c.}} 

\begin{document}

\begin{titlepage}

\vspace*{-2cm}
\begin{flushright}
LPT Orsay 16-64 \\
CFTP/16-013 \\
\vspace*{2mm}
\end{flushright}

\begin{center}
\vspace*{15mm}

\vspace{1cm}
{\LARGE \bf
Higgs EFT for 2HDM and beyond
} 
\vspace{1.4cm}

\renewcommand{\thefootnote}{\fnsymbol{footnote}}
{\bf Herm\`{e}s B\'{e}lusca-Ma\"{i}to$^{a}$, Adam Falkowski$^a$, Duarte Fontes$^b$,  \\ Jorge C. Rom\~ao$^b$, and Jo\~{a}o P. Silva$^b$}
\renewcommand{\thefootnote}{\arabic{footnote}}
\setcounter{footnote}{0}

\vspace*{.5cm}
\centerline{$^a${\it Laboratoire de Physique Th\'{e}orique, CNRS, Univ. Paris-Sud, Universit\'{e}  Paris-Saclay, 91405 Orsay, France}}
\centerline{$^b${\it CFTP, Departamento de F\'{i}sica, Instituto Superior T\'{e}cnico, Universidade de Lisboa,}}
\centerline{\it  Avenida Rovisco Pais 1, 1049 Lisboa, Portugal}

\vspace*{.2cm}

\end{center}

\vspace*{10mm}
\begin{abstract}\noindent\normalsize

We discuss the validity of the Standard Model  Effective Field Theory (SM EFT) as the low-energy effective theory for the two-Higgs-doublet Model (2HDM).
Using the up-to-date Higgs signal strength measurements at the LHC, one can  obtain a likelihood function for the Wilson coefficients of dimension-6 operators in the EFT Lagrangian.   
Given the matching between the 2HDM and the EFT, the constraints on the Wilson coefficients can be translated into  constraints on the parameters of the 2HDM Lagrangian.  
We discuss under which conditions such a procedure correctly reproduces the true limits on the 2HDM. 
Finally, we employ the SM EFT to identify  the pattern of the Higgs boson couplings that are needed to improve the fit to the current Higgs data.  
To this end, one needs, simultaneously, to increase the top Yukawa coupling, decrease the bottom Yukawa coupling, and induce a new contact interaction of the Higgs boson with gluons.  
We comment on how these modifications can be realized in the 2HDM extended by new colored particles.

\end{abstract}

\end{titlepage}
\newpage 

\renewcommand{\theequation}{\arabic{section}.\arabic{equation}} 

\section{Introduction}
\setcounter{equation}{0} 

Effective field theories (EFTs) allow one to describe the low-energy dynamics of a wide class of quantum theories \cite{Weinberg:1980wa,Manohar:1996cq,Kaplan:2005es}. 
The idea is to keep only the  subset of  light degrees of freedom, while discarding the heavy ones that cannot be produced on-shell in the relevant experimental setting. 
Virtual effects  of the heavy  particles on low-energy observables are represented by an infinite series of operators constructed out of the light fields. 

In the context of the LHC experiments, the light degrees of freedom are those of the Standard Model (SM), and the heavy ones correspond to hypothetical new particles. 
The low-energy effective description of such a framework is called the SM EFT, 
see e.g. \cite{Degrande:2012wf,Willenbrock:2014bja,Masso:2014xra,Pomarol:2014dya,Falkowski:2015fla,deFlorian:2016spz} for reviews.  
The SM EFT allows for a unified  description of many possible signals  of physics beyond the SM (BSM), assuming the new particles are  too heavy to be directly produced.
This model-independence is a great asset, given we currently have little clue about the more complete theory underlying the SM. 
Another  strength of this approach is that constraints on the EFT parameters can be easily translated into constraints on masses and couplings in specific BSM constructions. 
Thus, once experimental results are interpreted in the EFT language, there is no need to re-interpret them  in the context of  every possible model out there. 

A less appealing feature of EFTs  is that the Lagrangian  contains an infinite number of interaction terms and parameters, in contrast to renormalizable theories. 
In the SM EFT, these terms are organized in an expansion 
\begin{equation}
\label{eq:EFT_Lops}
{\cal L}_{\textrm{eff}}={\cal L}_{\rm SM}
+ \sum_{i}  \frac{c_{i}^{(6)}}{\Lambda^2} {\cal O}_{i}^{(6)} \
+ \sum_{i}  \frac{c_{i}^{(8)}}{\Lambda^4} {\cal O}_{i}^{(8)}  + \cdots\,,
\end{equation} 
where ${\cal L}_{\rm SM}$ is the SM Lagrangian,  $\Lambda$ is the mass scale of BSM physics, 
 each ${\cal O}_i^{(D)}$ is an $SU(3) \times SU(2) \times U(1)$  invariant operator of canonical  dimension $D$, and the parameters $c_i^{(D)}$  are called the Wilson  coefficients. 
Terms with odd $D$ are absent assuming baryon and lepton number conservation.

In practice, the series in \eref{EFT_Lops} must be truncated, such that one works with a finite set of parameters. 
In most  applications of the SM EFT,  terms  with $D\geq 8$ are neglected.
This corresponds to taking into account  the BSM effects  that scale as $\cO(m_W^2/\Lambda^{2})$, and neglecting those suppressed by higher powers of $\Lambda$. 
It is important to discuss the {\em validity} of such a procedure for a given experimental setting  \cite{Contino:2016jqw}. 
More precisely, the questions are 1) whether the truncated EFT gives a faithful description of the low-energy phenomenology of the underlying BSM model, and 2) to what extent  experimental constraints on the $D$=6 Wilson coefficients are affected by the neglected higher-dimensional operators.
Generically,  in the context of LHC Higgs studies the truncation is justified if $\Lambda$ is much larger than the electroweak scale. 
But, to address the validity issue more quantitatively and identify exceptional situations,  it is useful to turn to concrete models and compare the description on physical observables in the full BSM theory with that in the corresponding low-energy EFT. 
Such an  exercise provides valuable lessons about the validity range and limitations of the SM EFT.   
 
In this paper we perform that exercise for the $\mathbb{Z}_2$-symmetric CP-conserving two-Higgs-doublet model (2HDM). 
We compare the performance of the full model and its low-energy EFT truncated at $D$=6 to describe the Higgs signal strength measurements at the LHC.  
To this end, we first update the tree-level constraints on the 2HDM parameter space using the latest Higgs data from Run-1 and Run-2 of the LHC. 
We use the same  data to derive leading order  constraints on the parameters of the SM EFT.
Given the matching between the EFT and the 2HDM parameters \cite{Gorbahn:2015gxa,Brehmer:2015rna}, the EFT constraints can be subsequently recast as constraints on the parameter space of the 2HDM. 
By comparing the direct and the EFT approaches, we identify the validity range of the EFT framework where it provides an adequate description of the impact of 2HDM particles on the LHC Higgs data.

We also remark that neither the SM nor the 2HDM provides a very good fit to the Higgs data, mostly due to some tension with the measured rate of the $t \bar t h$ production  and $h \to b \bar b$ decays. 
If the current experimental hints of an enhanced $t \bar t h$ and suppressed $h \to b \bar b$ are confirmed by the future LHC data,  the 2HDM alone  will not be enough to explain these.   
Here the EFT approach proves to be very useful in suggesting extensions of the 2HDM that better fit the current Higgs data.
In particular, we show that a good fit requires simultaneous modifications of the EFT parameters controlling the top and bottom Yukawa couplings {\em and} the contact interaction of the Higgs boson with gluons. 
We show how these modifications can be realized in the 2HDM extended by new colored particles coupled to the Higgs.

This paper is organized as follows. 
In \sref{formalism} we review the 2HDM and its low-energy EFT. 
In \sref{comparison} we compare the direct and the EFT constraints on the parameter space imposed by  the Higgs measurements.   
In \sref{bthdm} we discuss how to improve  the fit to the LHC Higgs data by extending the 2HDM with new colored states coupled to the Higgs.

\section{Formalism}
\label{sec:formalism}
\setcounter{equation}{0} 

\subsection{CP-conserving 2HDM}

We start by reviewing the  (non-supersymmetric) 2HDM  \cite{Lee:1973iz,Gunion:1989we,Branco:2011iw}, closely following the formalism and notation of Ref.~\cite{Haber:2015pua}.
We consider two Higgs doublets $\Phi_1$ and $\Phi_2$, both transforming as $(1,2)_{1/2}$ under the SM gauge group. 
Both doublets may develop a vacuum expectation value (VEV) parametrized as $\langle \Phi_i^0 \rangle = \frac{v_i}{\sqrt{2}}$, with  
$v_1 = v \cos \beta \equiv v c_\beta$, $v_2 = v \sin \beta \equiv v s_\beta$, and $v = 246.2$~GeV.  
We assume that all  parameters in  the scalar potential are real, which implies the Higgs sector preserves the CP symmetry at the leading order. 

Furthermore, we assume that the Lagrangian is invariant under a discrete $\mathbb{Z}_2$ symmetry under which the doublets transform as $\Phi_1 \to +\Phi_1$ and $\Phi_2 \to - \Phi_2$. This symmetry is allowed to be broken only softly, that is to say, only by mass parameters in the Lagrangian. The $\mathbb{Z}_2$ symmetry  constrains the possible form of Yukawa interactions. There are four possible classes of 2HDM, depending on how the SM fermions transform under the $\mathbb{Z}_2$ symmetry. They are summarized in the following table:
	\begin{center}
		\begin{tabular}{ c | c | c | c | c }
			\hline\hline
			          & Type-I   & Type-II  & $\ell$-specific (Type-X) & Flipped (Type-Y) \\
			\hline
			Up-type   & $\Phi_2$ & $\Phi_2$ & $\Phi_2$                 & $\Phi_2$ \\
			Down-type & $\Phi_2$ & $\Phi_1$ & $\Phi_2$                 & $\Phi_1$ \\
			Leptons   & $\Phi_2$ & $\Phi_1$ & $\Phi_1$                 & $\Phi_2$ \\
			\hline\hline
		\end{tabular}
	\end{center}
It is often more convenient to work with  linear combinations  of $\Phi_1$ and $\Phi_2$ defined by the rotation  
	\begin{equation}
		\begin{pmatrix} H_1 \\ H_2 \end{pmatrix} = \begin{pmatrix}
			 c_\beta & s_\beta \\
			-s_\beta & c_\beta
		\end{pmatrix}
 \begin{pmatrix} \Phi_1 \\ \Phi_2 \end{pmatrix}. 
	\end{equation}
It follows that  $\langle H_1^0 \rangle = \frac{v}{\sqrt{2}}$, $\langle H_2^0 \rangle =0$.
Note that $H_1$ and $H_2$, unlike $\Phi_i$, are not eigenstates of the $\mathbb{Z}_2$  symmetry. 
The linear combinations  $H_i$ define the so-called {\em Higgs basis} \cite{Botella:1994cs}, while the original doublet $\Phi_i$ are referred to as the  $\mathbb{Z}_2$ basis. 

In the Higgs basis, the scalar potential takes the form:
	\begin{equation}\begin{split}
		V(H_1, H_2) =\;& Y_1 |H_1|^2 + Y_2 |H_2|^2 + (Y_3 H_1^\dagger H_2 + \hc) + \frac{Z_1}{2} |H_1|^4 + \frac{Z_2}{2} |H_2|^4 \\
			& + Z_3 |H_1|^2 |H_2|^2 + Z_4 (H_1^\dagger H_2)(H_2^\dagger H_1) \\
			& + \left\{ \frac{Z_5}{2} (H_1^\dagger H_2)^2 + (Z_6 |H_1|^2 + Z_7 |H_2|^2)(H_1^\dagger H_2) + \hc \right\},
	\end{split}\end{equation}
	where the parameters $Y_i$ and $Z_i$ are all real.
The $\mathbb{Z}_2$ symmetry is manifested by the fact  that only 5 of the $Z_i$ are independent, as they satisfy 2 relations: 
	\begin{equation}\begin{aligned}
		    Z_2 - Z_1 &= \frac{1 - 2 s_\beta^2}{s_\beta c_\beta}  (Z_6 + Z_7) \, , \\ 
		Z_{345} - Z_1 &= \frac{1 - 2 s_\beta^2}{s_\beta c_\beta} Z_6 - \frac{2 s_\beta c_\beta}{1 - 2 s_\beta^2} (Z_6 - Z_7) \, ,
	\end{aligned}\end{equation}
	where $Z_{345} \equiv Z_3 + Z_4 + Z_5$. 
The Yukawa couplings are given by:
	\begin{equation}\begin{split}
		\mathcal{L}_\text{Yukawa} =&\;
			    - \tilde H_1^\dagger \overline{u_R} Y_u q_L - H_1^\dagger \overline{d_R} Y_d q_L - H_1^\dagger \overline{e_R} Y_e \ell_L \\
			&\; - \frac{\eta_u}{\tan\beta} \tilde H_2^\dagger \overline{u_R} Y_u q_L - \frac{\eta_d}{ \tan\beta} H_2^\dagger \overline{d_R} Y_d q_L - \frac{\eta_e}{\tan\beta} H_2^\dagger \overline{e_R} Y_e \ell_L  + \hc 
			\quad , 
	\end{split}\end{equation}
	where $\tilde H_i = i \sigma_2 H_i^*$, and the coefficients of the $H_2$ Yukawa couplings are summarized in the table below: 
	\begin{center}
		\begin{tabular}{ c | c | c | c | c }
			\hline\hline
			         & Type-I & Type-II      & Type-X & Type-Y  \\
			\hline
			$\eta_u$ & $1$    & $1$          & $1$          & $1$          \\
			$\eta_d$ & $1$    & $-\tan^2\beta$ & $1$          & $-\tan^2\beta$ \\
			$\eta_e$ & $1$    & $-\tan^2\beta$ & $-\tan^2\beta$ & $1$          \\
			\hline\hline
		\end{tabular}
	\end{center}	

In the Higgs basis, the doublets can be parametrized as
	\begin{equation}
		H_1 = \begin{pmatrix} -\imath G^+ \\ \frac{1}{\sqrt{2}} (v + s_{\beta-\alpha} h   + c_{\beta-\alpha} H_0 + \imath G_z) \end{pmatrix} \;, \;
		H_2 = \begin{pmatrix} H^+ \\ \frac{1}{\sqrt{2}} (c_{\beta-\alpha} h  -  s_{\beta-\alpha} H_0 + \imath A) \end{pmatrix},
	\end{equation}
	where $G^\pm$ and $G_z$ are the Goldstone bosons eaten by $W^\pm$ and $Z$, while $H^\pm$ and $A$ are the charged scalar and neutral pseudo-scalar eigenstates. 
The two neutral scalars $h$, $H_0$ are mass eigenstates, while the parameter $c_{\beta - \alpha} \equiv \cos (\beta - \alpha)$ determines their embedding in the two doublets $H_i$.\footnote{The angle $\alpha$ can be defined as the rotation angle connecting the components of the original Higgs doublets $\Phi_1$ and $\Phi_2$ to the mass eigenstates.}
In the following we will identify $h$ with the 125 GeV Higgs boson. 
	
The equations of motion for $H_1$ and $H_2$ imply the vacuum relations:
	\begin{equation}
		Y_1 = - \frac{Z_1}{2} v^2 \quad , \quad Y_3 = - \frac{Z_6}{2} v^2 \, .
	\end{equation}
The masses of the charged scalar and the pseudo-scalar are given by:
	\begin{equation}
	\label{eq:mhp}
		m_{H^+}^2 = Y_2 + \frac{Z_3}{2} v^2 \quad , \quad m_A^2 = Y_2 + \frac{Z_3 + Z_4 - Z_5}{2} v^2 \, .
	\end{equation}
	The mixing angle is related to the parameters of the potential by:
	\begin{equation}
		\label{eq:bma}
		\frac{1}{2} \tan(2(\beta - \alpha)) \equiv - \frac{s_{\beta-\alpha} c_{\beta-\alpha}}{1 - 2 c_{\beta-\alpha}^2} =
		 \frac{Z_6}{\frac{Y_2}{v^2} + Z_{345}/2 - Z_1} \, .
	\end{equation}
	The masses of the neutral scalars can be written as:
	\begin{equation}\begin{aligned}
		\label{eq:mhn}
		m_h^2 &= v^2 \left( Z_1 + \frac{c_{\beta-\alpha}}{s_{\beta-\alpha}} Z_6 \right)
		\\
		m_{H_0}^2 & 
		= \frac{s_{\beta-\alpha}^2 Y_2 + Z_{345} s_{\beta-\alpha}^2 v^2/2 - Z_1 c_{\beta-\alpha}^2 v^2}{1 - 2 c_{\beta-\alpha}^2} . 
	\end{aligned}\end{equation}
Finally, the couplings of the CP-even scalar $h$, 
 to the electroweak gauge bosons are given by 
\beq
\cL_{hVV}  ={h \over v} \left (2 m_W^2 W_\mu^+ W^{\mu,-} + m_Z^2 Z_\mu Z^\mu \right)  \sqrt{1-  c_{\beta-\alpha}^2}  , 
\eeq 
and to the fermions by
\beq
\label{eq:lhff}
\cL_{hff}  =  - {h \over v} \sum_f m_f  \bar f f \left (  \sqrt{1-  c_{\beta-\alpha}^2} + \eta_f {c_{\beta-\alpha} \over \tan \beta} \right ).
\eeq 
By convention, the sign of the $h$ couplings to $WW$ and $ZZ$ is fixed to be positive (this can always be achieved, without loss of generality, by redefining the Higgs boson field as $h \to - h$). 
On the other hand,  the sign of the $h$ couplings to fermion may be positive or negative, depending on the value of 
$c_{\beta-\alpha}$ and  $\tan \beta$. 
The \emph{alignment limit} is defined by $c_{\beta-\alpha} \to 0$, that is to say, when $h$ has SM couplings.
There is a strong evidence, both from Higgs and from electroweak precision measurements, that the couplings of the 125 GeV boson to $W$ and $Z$ bosons are very close to those predicted by the SM. 
Therefore the 2HDM has to be near the alignment limit to be phenomenologically viable. 
From Eq.~(\ref{eq:bma}), the condition for alignment is: 
\beq 
\label{eq:alignmentcondition}
|Z_6| \ll | Y_2/v^2 +Z_{345}/2 - Z_1|.
\eeq 
One way to satisfy this is by making $Y_2$ large,  $Y_2 \gg v^2$, which is called the \emph{decoupling limit} because then $H_0$, $A$ and $H^+$ become heavy.
 Another way to ensure alignment is to take $|Z_6|$ small enough, $|Z_6| \ll 1$. 
 If the condition \eref{alignmentcondition} is satisfied with $Y_2 \lesssim v^2$ then we speak of \emph{alignment without decoupling}. 

\subsection{Low-energy EFT}
\label{sec:eftmatching}

For $Y_2 \equiv \Lambda^2  \gg v^2$ and $Y_1 \sim Y_3 \sim v$,  \eref{mhp} and \eref{mhn} imply $m_A \sim m_{H_+} \sim m_{H_0} \sim  \Lambda$, and  the spectrum below the scale $v$ is that of the SM. 
Consequently,  we can describe Higgs production and decays at the LHC in the framework of the so-called SM EFT, 
where the heavy particles are integrated out, and their effects are represented by operators with canonical dimensions $D>4$ added to the SM. 
Below we discuss the Lagrangian of the low-energy effective theory for the 2HDM, treating $1/\Lambda$ as the expansion parameter.  
We first review the known results concerning the $D$=6 operators in the EFT with tree-level matching \cite{Gorbahn:2015gxa,Brehmer:2015rna}. 
This is enough for the purpose of this paper, in which the main focus is the accuracy of the EFT to describe the current LHC Higgs measurements.    
Matching beyond $D$=6 and tree level was discussed in Refs.~\cite{Henning:2014wua,Gorbahn:2015gxa}, and we will come back to it in an upcoming publication \cite{ustoappear}. 

The simplest way to derive the  tree-level matching is  by integrating out the field $H_2$ and identifying $H_1$ with the SM Higgs doublet.
The procedure is to: 1) solve the \emph{linearized} equations of motion for $H_2$ as a function of the light fields (the scalar doublet $H_1$, fermions, and gauge fields), and 2) insert the solution in the original Lagrangian. Furthermore, restricting to $D=6$ operators in the EFT, one can ignore all derivative terms in the $H_2$ equation of motion. The linearized equation of motion for $H_2$ with derivative terms dropped is solved as:
	\begin{equation}
	\label{eq:2hdm_sold6}
		\Lambda^2 H_2 \approx - H_1  \left[ Y_3  + Z_6 H_1^\dagger H_1 \right] - \frac{\eta_f}{ \tan\beta} \bar f_R Y_f f_L \, .
	\end{equation}
	Plugging this back, renaming $H_1 \to H$,  and keeping terms up to $1/\Lambda^2$,  the effective Lagrangian takes the form:
	\begin{equation}
	\label{eq:2hdm_leff}
		\mathcal{L}_\text{eff} = \mathcal{L}_{SM} +
		\frac{1}{\Lambda^2} \left[ Z_6 H^\dagger \left( H^\dagger H + Y_3 \right) 
		    + \frac{\eta_f}{ \tan\beta} \bar f_R Y_f f_L \right]
		\left[ Z_6 H \left ( H^\dagger H+ Y_3 \right) + \frac{\eta_f}{ \tan\beta} \bar f_L Y_f f_R \right] \, .
	\end{equation}
The terms proportional to $Y_3$ can be absorbed in a re-definition of the SM parameters, and they do not have observable consequences. 
On the other hand, the genuine $D$=6 terms in \eref{2hdm_leff} are in principle observable. 	
We are interested in the impact of these $D$=6 operators on the Higgs boson couplings probed at the LHC. 
Quite generally, in the SM EFT with $D$=6 operators the CP-conserving Higgs boson couplings to two SM fields can be parametrized as
\cite{deFlorian:2016spz,Falkowski:2001958}:  
\begin{eqnarray}
\label{eq:PEL_lh}
 {\cal L}_{\rm h} &= & {h \over v} \left [ 
  \left (1 +  \delta c_w \right )  {g_L^2 v^2 \over 2} W_\mu^+ W^{\mu,-} 
  +    \left (1 +  \delta c_z \right )  {(g_L^2+g_Y^2) v^2 \over 4} Z_\mu Z^\mu
\right . \nonumber \\ & & \left . 
+ c_{ww}  {g_L^2 \over  2} W_{\mu \nu}^+  W^{\mu\nu,-}  
+ c_{w \Box} g_L^2 \left (W_\mu^- \partial_\nu W^{\mu \nu,+} + {\mathrm h.c.} \right )  
+c_{z \Box} g_L^2 Z_\mu \partial_\nu Z^{\mu \nu} + c_{\gamma \Box} g_L g_Y Z_\mu \partial_\nu A^{\mu \nu}
\right . \nonumber \\ & & \left . 
+  c_{gg} {g_s^2 \over 4 } G_{\mu \nu}^a G^{\mu \nu,a}   + c_{\gamma \gamma} {e^2 \over 4} A_{\mu \nu} A^{\mu \nu} 
+ c_{z \gamma} {e \sqrt{g_L^2 + g_Y^2}  \over  2} Z_{\mu \nu} A^{\mu\nu} + c_{zz} {g_L^2 + g_Y^2 \over  4} Z_{\mu \nu} Z^{\mu\nu}
\right . \nonumber \\ & & \left .  
- \sum_{f \in u,d,e} \sum_{ij}    
\sqrt{m_{f_i} m_{f_j}} \left ( \delta_{ij} + [\delta y_f]_{ij}   \right ) \bar f_{R,i} f_{L,j} + {\rm h.c.},
\right ]  - (\lambda +  \delta \lambda_3)  v h^3 .  
 \end{eqnarray} 
The effect of $D$=6 operators  	in \eref{2hdm_leff} is to shift the Higgs couplings to the SM fermions and to itself: 
	\begin{equation}\begin{gathered}
	\label{eq:map_tree}
		[\delta y_{u,d,e}]_{ij} = - \frac{\eta_{u,d,e}}{\tan \beta} Z_6 \frac{v^2}{\Lambda^2} \delta_{ij}
		\quad, \quad
		\delta\lambda_3 = - \frac{3 Z_6^2}{2} \frac{v^2}{\Lambda^2}
		\, .
	\end{gathered}\end{equation}
On the other hand, \emph{at tree-level and restricting to dimension-6 operators in the EFT}, there are no corrections to the Higgs boson interactions with gauge bosons:
	\begin{equation}
		\delta c_w  = \delta c_z =  c_{ww} = c_{zz} = c_{\gamma\gamma} = c_{z\gamma} = c_{gg} = c_{z\Box} = c_{w\Box} = c_{\gamma \Box} =  0 \, .
	\end{equation} 

	One can check that the couplings of the Higgs in the effective theory described by the Lagrangian in \eref{2hdm_leff} are the same as the couplings of $h$ in the 2HDM \emph{expanded to linear order in } $c_{\beta-\alpha}$, once we identify:
	\begin{equation}
		c_{\beta-\alpha} \leftrightarrow - \frac{Z_6 v^2}{\Lambda^2} \, .
	\end{equation}
	This identification is consistent with \eref{bma} when $Y_2 \gg v^2$.

We also comment on the interesting case of alignment without decoupling. Our EFT is supposed to be a good description of the 2HDM in the decoupling limit where all the additional scalars are heavy. In general, the EFT will not work when one or more scalars are light, even in the alignment limit. Indeed, if one of the new Higgs scalars are light, $2 \to 2$ fermion scattering will display a pole at the energy equals to the scalar's mass, which cannot be captured by the 4-fermion operators in \eref{2hdm_leff}. Similarly, double Higgs production will have a pole at the new mass (if the other neutral scalar is light), which again cannot be described by the operators \eref{2hdm_leff}.

However, it is possible that certain low-energy observables can still be adequately described by our EFT, even when the 2HDM has additional light scalars with mass of order  $m_h$. The Higgs couplings to matter are such observables, provided the 2HDM is in the alignment limit. 
More precisely, from the constraints on the couplings $\delta y_f$ one can correctly infer constraints on the parameters of the 2HDM in the limit of alignment without decoupling. However, to this end, the mapping between the parameters of the EFT and the 2HDM has to be modified: instead of \eref{map_tree}, we have to use the following map:
	\begin{equation}\begin{aligned}
	\label{eq:map_alignment}
		[\delta y_{u,d,e}]_{ij} &= - \frac{\eta_{u,d,e}}{ \tan\beta} Z_6 \frac{v^2}{Y_2 + \frac{v^2}{2} (Z_{345} - 2 Z_1)} \delta_{ij} \, .  
	\end{aligned}\end{equation}
	This formula follows from expanding the 2HDM expressions for the Yukawa couplings of $h$ to the leading order in  $c_{\beta-\alpha}$. Using that, \eref{map_alignment} can be obtained by expanding the 2HDM Higgs couplings in $c_{\beta-\alpha}$ and using the expression for $c_{\beta-\alpha}$ that is also valid for alignment without decoupling. 
	The new terms in the matching formulas are negligible in the decoupling limit $Y_2 \gg v^2$, in which case they are higher order in  the $v^2/\Lambda^2$ expansion.  
	However they can be very important in the case of alignment without decoupling when $Y_2 \lesssim v^2$.
Such a way of extending the validity range of the EFT by adding  higher order terms in the matching formula is similar to $v$-improved matching advocated in Ref.~\cite{Brehmer:2015rna,Freitas:2016iwx}.

\section{Comparison of EFT and 2HDM descriptions of Higgs couplings}
\label{sec:comparison}

\begin{table}[tb!]
\begin{center}
\renewcommand*{\arraystretch}{1.1} 
\begin{tabular}{|c|c|c|c|c|}
\hline
Channel   & Production   &   Run-1 &  ATLAS Run-2  &  CMS Run-2 \\   
\hline
$\gamma \gamma$  & $ggh$ & $1.10^{+0.23}_{-0.22} $ & $0.62^{+0.30}_{-0.29}$  \cite{ATLAS:2016hru} &  $0.77^{+0.25}_{-0.23}$   \cite{CMS:2016ixj}
\\ \cline{2-5} 
         & VBF  &  $1.3^{+0.5}_{-0.5}$ & $2.25^{+0.75}_{-0.75}$ \cite{ATLAS:2016hru}&  $1.61^{+0.90}_{-0.80}$    \cite{CMS:2016ixj}
\\ \cline{2-5} 
         & $Wh$  &  $0.5^{+1.3}_{-1.2}$ & - & -
\\ \cline{2-5}
 	& $Zh$  &   $0.5^{+3.0}_{-2.5}$ & - & - 
\\ \cline{2-5} 
& $Vh$  &  - & $0.30^{+1.21}_{-1.12}$  \cite{ATLAS:2016hru}    &  - 	  
\\ \cline{2-5} 
& $t \bar t h$ & $2.2^{+1.6}_{-1.3}$& $-0.22^{+1.26}_{-0.99}$  \cite{ATLAS:2016hru} &   $1.9^{+1.5}_{-1.2}$  \cite{CMS:2016ixj}
\\ \hline 
$Z \gamma$ & incl. & $1.4^{+3.3}_{-3.2}$   & - & - 
\\ \hline 
$Z Z^*$   & $ggh$ & $1.13^{+0.34}_{-0.31} $   & $1.34^{+0.39}_{-0.33} $  \cite{ATLAS:2016hru} &  $0.96^{+0.40}_{-0.33}$ \cite{CMS:2016ilx}
\\ \cline{2-5} 
         & VBF   & $0.1^{+1.1}_{-0.6}$ &  $3.8^{+2.8}_{-2.2}$    \cite{ATLAS:2016hru}  &  $0.67^{+1.61}_{-0.67}$  \cite{CMS:2016ilx}
\\ \hline 
$WW^*$   & $ggh$ & $0.84^{+0.17}_{-0.17} $  & - & - 
\\ \cline{2-5} 
         & VBF  &  $1.2^{+0.4}_{-0.4}$ & - & - 
\\ \cline{2-5}  
         & $Wh$  &  $1.6^{+1.2}_{-1.0}$ & - & - 
\\ \cline{2-5} 
 	& $Zh$  &   $5.9^{+2.6}_{-2.2}$  & - & - 
\\ \cline{2-5} 
& $t \bar t h$ & $5.0^{+1.8}_{-1.7}$  & - & -  
\\ \cline{2-5}  
& incl.  & - &	-	&  $0.3 \pm 0.5$   \cite{CMS:2016nfx}
\\ \hline 
$\tau^+ \tau^-$   & $ggh$ & $1.0^{+0.6}_{-0.6} $  & - & -  
\\ \cline{2-5} 
         & VBF  &  $1.3^{+0.4}_{-0.4}$ & - & -  
\\ \cline{2-5} 
         & $Wh$  &  $-1.4^{+1.4}_{-1.4}$ & - & -  
\\ \cline{2-5} 
 	& $Zh$  &   $2.2^{+2.2}_{-1.8}$   & - & -  
\\ \cline{2-5} 
& $t \bar t h$ & $-1.9^{+3.7}_{-3.3}$ & - & -  
\\ \hline 
$b \bar b$ & VBF  &-   & $-3.9^{+2.8}_{-2.9}$ \cite{ATLAS:2016lgh} &  $-3.7^{+2.4}_{-2.5}$  \cite{CMS:2016mmc}
\\ \cline{2-5} 
& $Wh$ & $1.0^{+0.5}_{-0.5}$  & - & -  
\\ \cline{2-5} 
  &	$Zh$ & 	$0.4^{+0.4}_{-0.4}$	 & - & -  
  \\ \cline{2-5}  
    &	$Vh$ & -  & $0.21^{+0.51}_{-0.50}$  \cite{ATLAS:2016pkl}  &  - 
  \\ \cline{2-5}  
  &	 $t \bar t h$ & 	$1.15^{+0.99}_{-0.94}$	   & $2.1^{+1.0}_{-0.9}$ \cite{ATLAS:2016awy} &  $-2.0^{+1.8}_{-1.8}$ \cite{CMS:2016qwm}
\\ \hline 
 $\mu^+ \mu^-$ & incl. & $0.1^{+2.5}_{-2.5}$ & $-0.8^{+2.2}_{-2.2}$ \cite{ATLAS:2016zzs} & -  
\\ \hline 
 multi-$\ell$ & cats.  & - & $2.5^{+1.3}_{-1.1}$ \cite{ATLAS:2016ldo} 	& $2.3^{+0.9}_{-0.8}$  \cite{CMS:2016vqb}
\\ \hline 
\end{tabular}
\quad 
\end{center}
\caption{
\label{tab:HIGGS_datarun12}
The Higgs signal strength in various channels measured at the LHC. 
For the Run-1, the $Z \gamma$ signal strength is a naive Gaussian  combination of   ATLAS  \cite{Aad:2015gba}  and CMS \cite{Chatrchyan:2013vaa}  results, 
and all the remaining numbers are taken from the ATLAS+CMS combination paper \cite{Khachatryan:2016vau}.
Correlations between different Run-1 measurements quoted in Fig.~27 of \cite{Khachatryan:2016vau} are taken into account.
}
\end{table}

In this section we discuss constraints from the Higgs signal strength measurements at the LHC. 
To this end, we use the results summarized in \tref{HIGGS_datarun12} which also include preliminary Run-2 results.  
First we update the tree-level constraints on the $c_{\beta-\alpha}$-$\tan \beta$ plane of the various $\mathbb{Z}_2$ -symmetric versions of the 2HDM. 
The same LHC data can also be used to derive leading order  constraints on the parameters of the SM EFT with  $D$=6 operators. 
These can be subsequently recast as constraints on the 2HDM parameters using the tree-level matching in \sref{eftmatching}. 
As long as the extra scalars of the 2HDM are heavy, we expect that the EFT should give an adequate description of the Higgs physics, and then the constraints should be the same regardless whether we obtain them directly or via the EFT. 
The goal of this section is to validate this expectation and quantify the validity range of the EFT for the 2HDM. 
Finally, we will also compare the results obtained by the above analyses with more sophisticated parameter scans of the 2HDM, that take into account the limits from precision measurements, unitarity, and boundedness of the Higgs potential.

\subsection{Update of Higgs constraints on 2HDM}
\label{sec:2hdmfits}

\begin{figure}[tb]
\includegraphics[width=0.45 \textwidth]{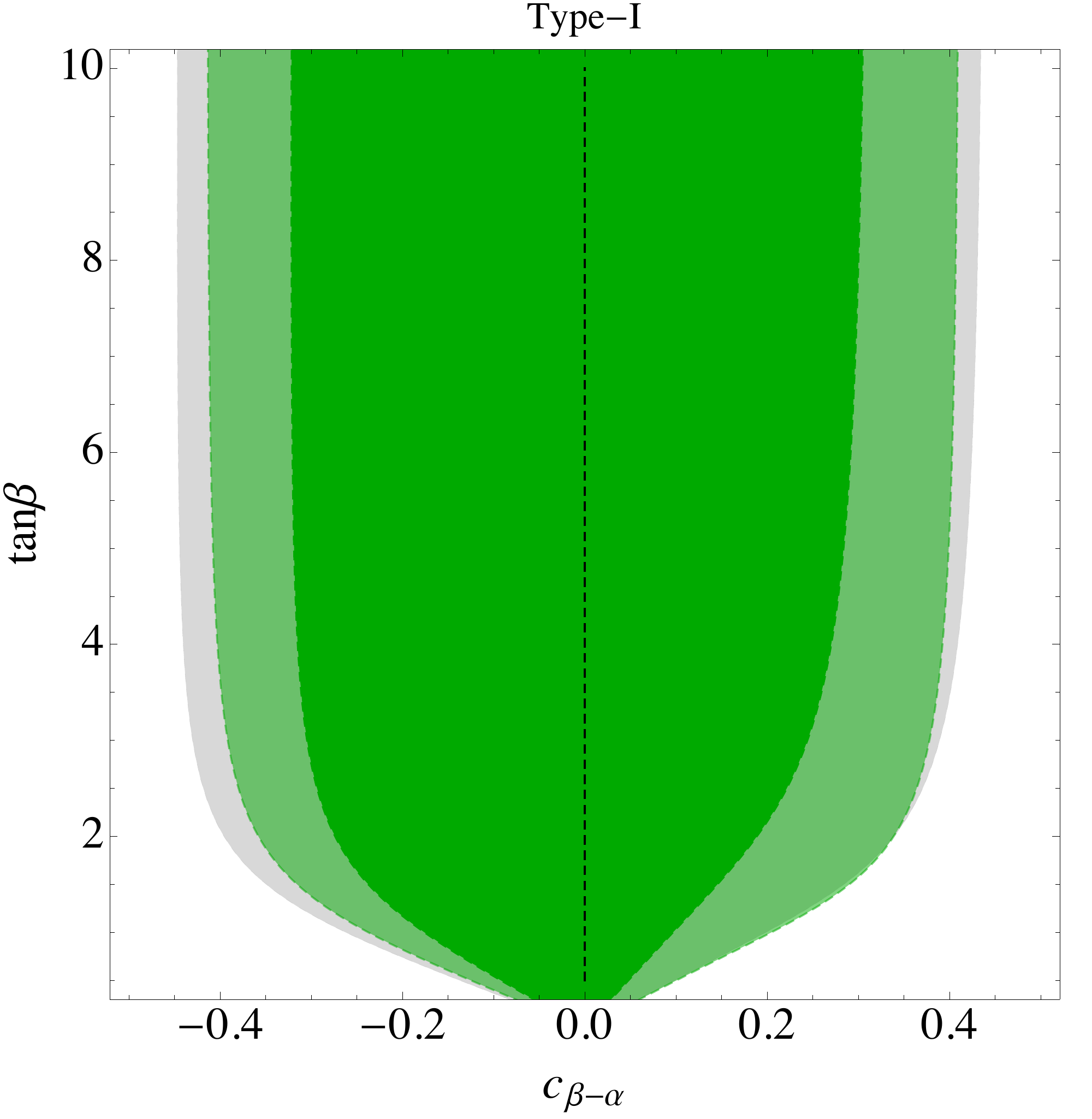}
\quad
\includegraphics[width=0.45 \textwidth]{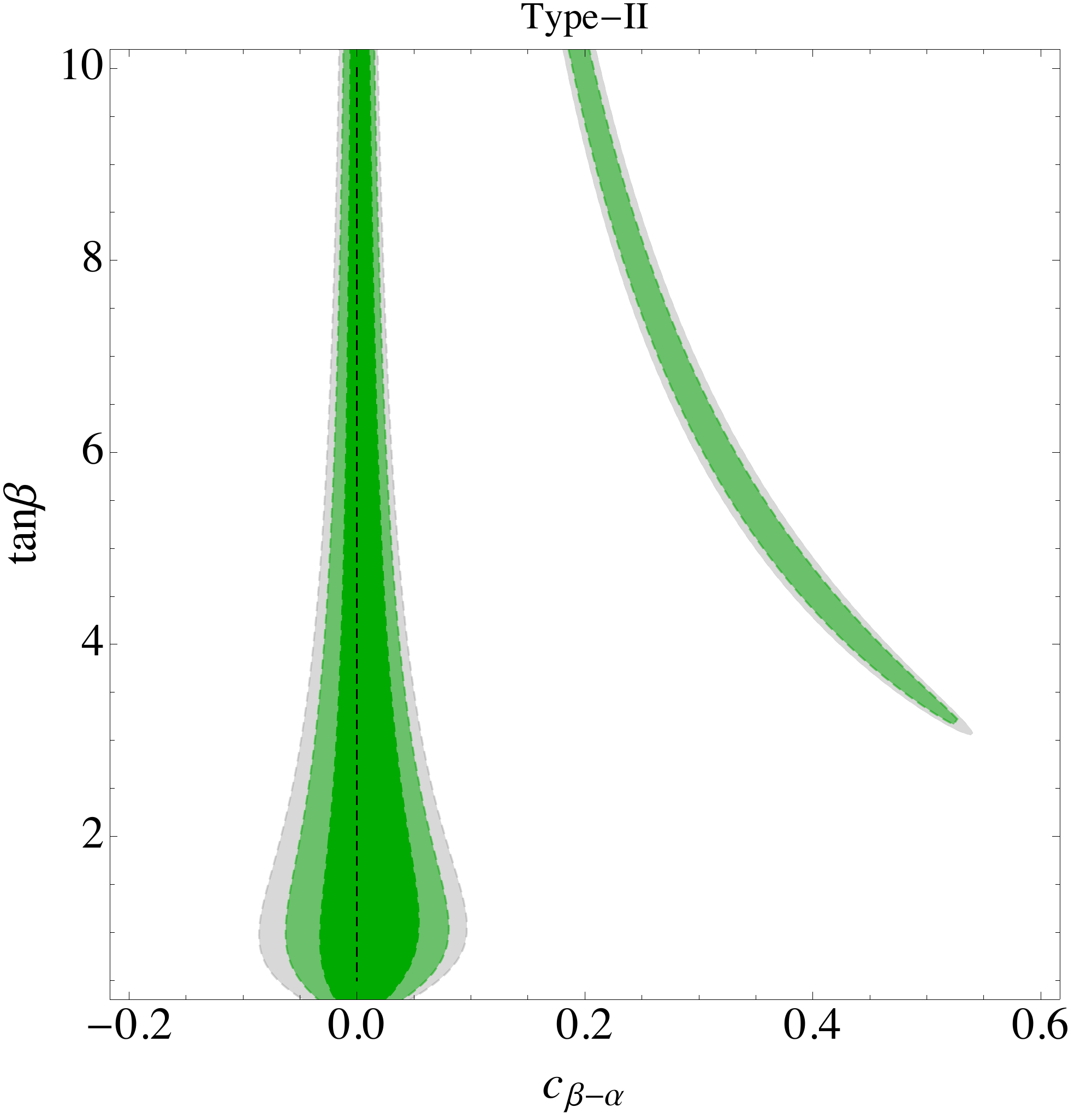}
\caption{Constraints from the LHC Higgs data on the parameter space of the type-I (left) and type-II (right)  2HDM. 
We show the 68\% (darker green) and 95\% (lighter green) CL region in the $c_{\beta-\alpha}$-$\tan \beta$ plane. 
The gray area  is the  95\% CL region obtained using Higgs Run-1 data only. 
}
\label{fig:THDM_1}
\end{figure} 

\begin{figure}[tb]
\includegraphics[width=0.45 \textwidth]{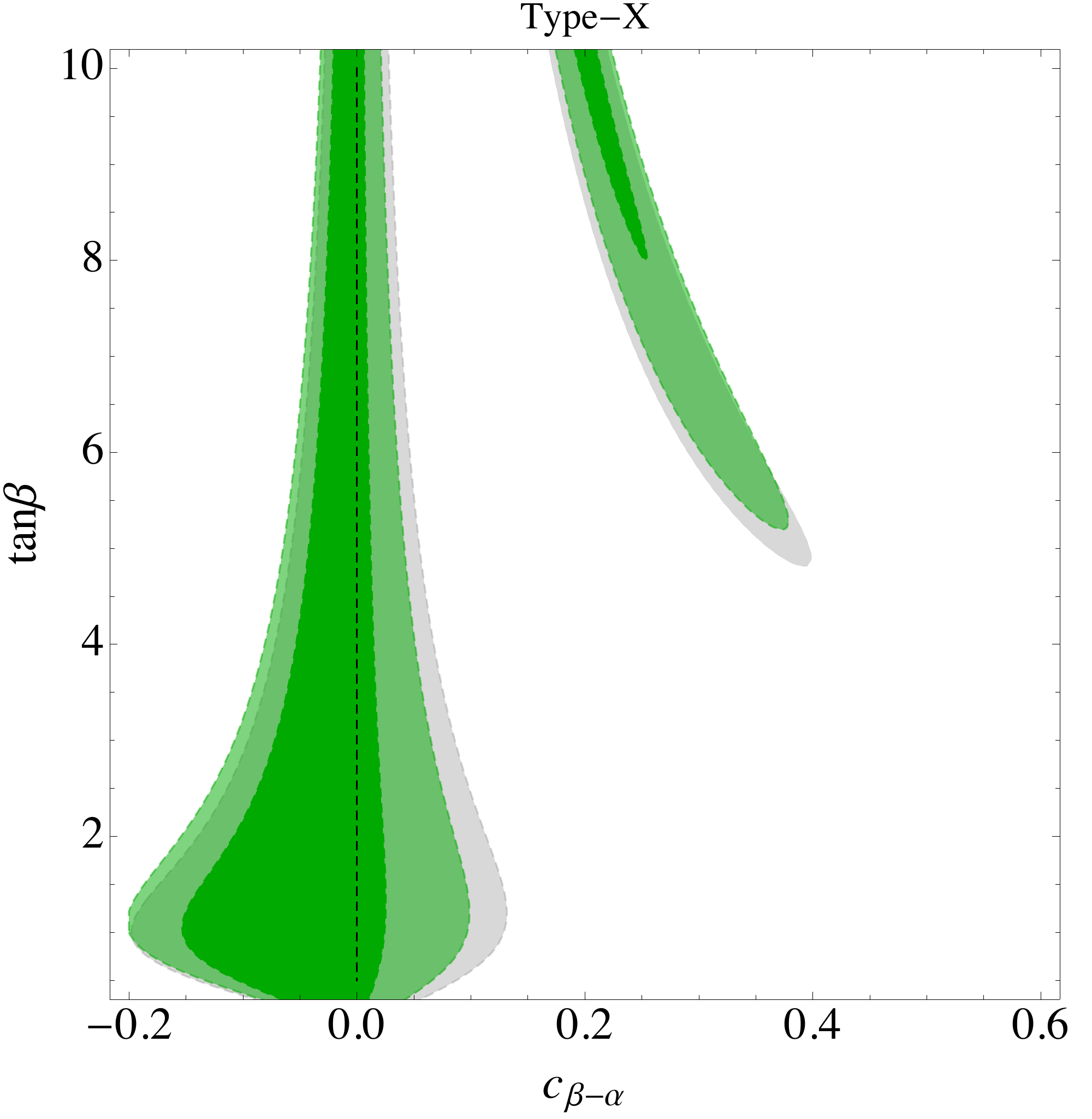}
\quad
\includegraphics[width=0.45 \textwidth]{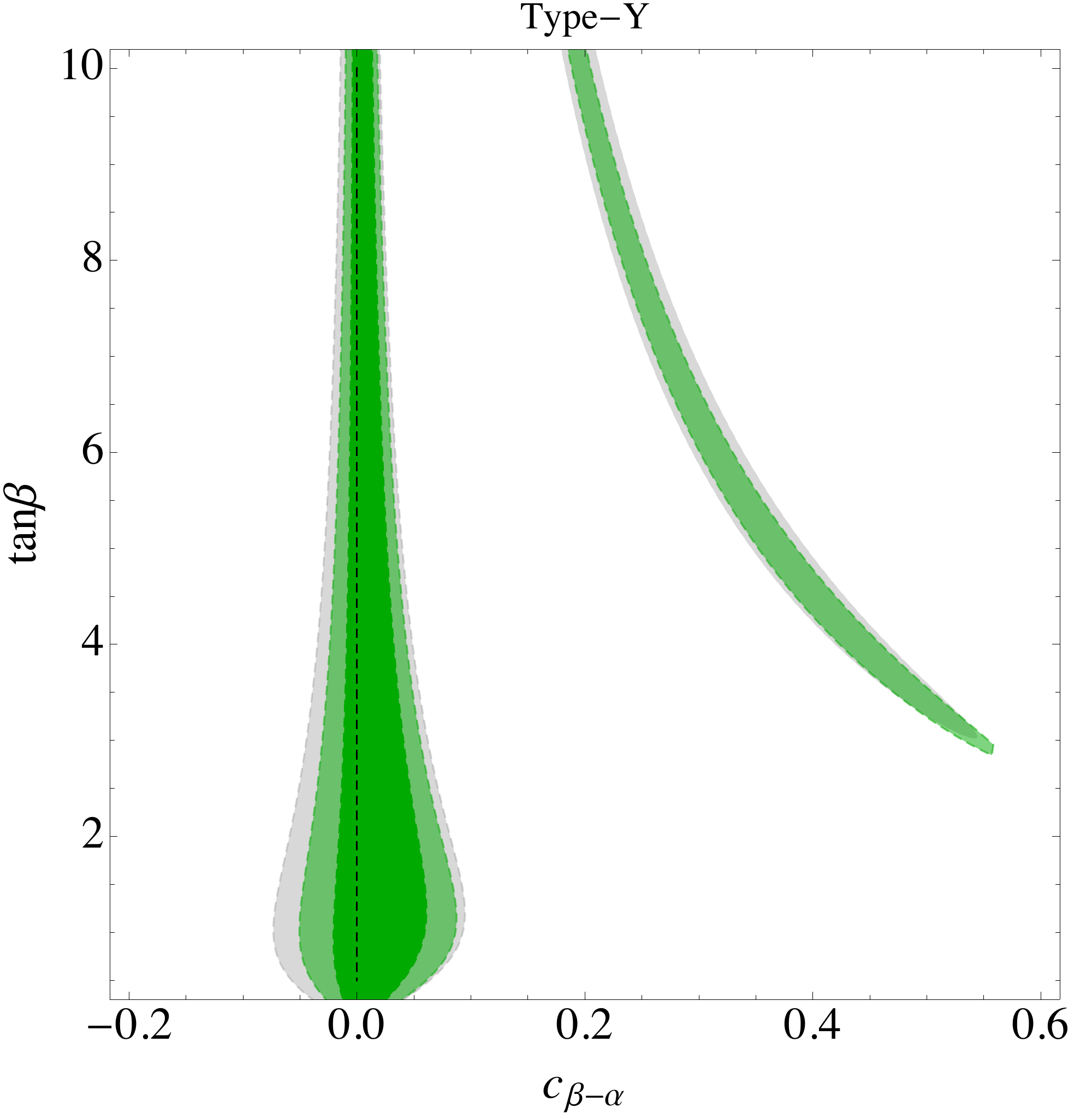}
\caption{Constraints from LHC Higgs data on the parameter space of the type-X (left) and  type-Y (right)  2HDM. 
We show the 68\% (darker green) and 95\% (lighter green) CL region in the $c_{\beta-\alpha}$-$\tan \beta$ plane. 
The gray area  is the  95\% CL region obtained using Higgs Run-1 data only. 
}
\label{fig:THDM_2}
\end{figure} 

We first show the constraints on various 2HDM scenarios  from the LHC studies of the 125 GeV Higgs. 
The results are shown in \fref{THDM_1} and  \fref{THDM_2}.  
The constraints are the weakest in the type-I model, especially for large $\tan \beta$. 
The reason is that, in this case, corrections to fermionic Higgs couplings are suppressed by $\tan \beta$ for a fixed $c_{\beta - \alpha}$.  
Although corrections to the Higgs couplings to $W$ and $Z$ do not have that suppression, 
they appear only at the {\em quadratic} order in $c_{\beta - \alpha}$ and therefore they become significant only for large $|c_{\beta - \alpha}|$. 
As a result, in the type-I model the 125 GeV Higgs boson can be further away from the alignment limit, with the modest bound  $|c_{\beta - \alpha}| \lesssim 0.4$ for large enough $\tan \beta$.

The constraints on $c_{\beta - \alpha}$ are much more stringent for the type-II, type-X, and type-Y  scenarios. 
In those cases, for a fixed $c_{\beta - \alpha}$, there is always a modification of some fermionic Higgs coupling  that is not suppressed by $\tan \beta$. 
In the generic region of the parameter space the bound is  $|c_{\beta - \alpha}| \lesssim 0.1$-$0.2$ for $\tan \beta \sim 1$, and even more stringent for smaller and larger $\tan \beta$. 
These  scenarios also display a separate region of the parameter space where a large $c_{\beta - \alpha}$ is allowed. 
It corresponds to the situation when corrections to the down-type quark and/or lepton Yukawa couplings  flip their sign but leave the absolute values close to the SM one \cite{Ferreira:2014naa,Fontes:2014tga,Ferreira:2014sld}.
Note that 3 distinct situations can arise: when the down-type Yukawas become negative (type-Y), when the lepton Yukawas become negative (type-X), and when both become negative (type-II). 
We refer to all these 3 cases  as the ``wrong-sign Yukawa" region.  
The Higgs observables  are currently weakly sensitive to the sign of the bottom and tau Yukawa, 
therefore these somewhat fine-tuned regions remain consistent with the data.  
Future precision tests may resolve the sign of the bottom Yukawa \cite{Ginzburg:2001ss,Carmi:2012yp,Chiang:2013ixa,Ferreira:2014naa,Fontes:2014tga,Ferreira:2014dya,Modak:2016cdm}, but that may be challenging for the tau Yukawa.

The qualitative shape of the favored regions is the same as that obtained from Run-1 Higgs data.
The effect of the preliminary Run-2 data is to make the constraints somewhat more stringent.

\subsection{Higgs constraints on EFT}
\label{sec:eftfits}

As explained in \sref{eftmatching}, at the leading order in the $1/\Lambda^2$ expansion the 2HDM induces corrections only to Higgs Yukawa- and self-interactions.
The latter can be  probed by non-resonant double Higgs production but,  given the current level of precision, the existing limits on the Higgs cubic self-coupling do not lead to any interesting constraints on the 2HDM parameter space. 
Therefore,at order $1/\Lambda^2$,  the parameters of the SM EFT relevant for the 2HDM are  the three $\delta y_f$ characterizing corrections to the SM Higgs Yukawa couplings, see \eref{PEL_lh}.  
As an intermediate step in connecting the 2HDM to the SM EFT, we can derive the constraints on these 3 EFT parameters. 
We find that the Run-1 and Run-2 Higgs data lead to the following constraints:
\beq
\label{eq:3par}
\bvec \delta y_ u \\ \delta y_ d \\\delta y_ e  \evec  = 
\bvec -0.11 \pm 0.10 \\ -0.14 \pm 0.11  \\ 0.02 \pm 0.13 \evec, 
\quad \rho =  \left ( \ba{ccc} 
1 & 0.83 & 0.25  \\ 
0.83 & 1 & 0.29  \\ 
0.25 &  0.29 & 1 \ea \right ),  
\eeq 
where the quoted  uncertainties correspond to 1~$\sigma$, and $\rho$ is the correlation matrix. 
The central values are close to the SM point, with $\chi^2_{\rm SM} - \chi^2_{\rm min} \approx 1.8$. 
These results are obtained by expanding the EFT predictions for the Higgs signal strength observables to the linear order in $\delta y_f$, and ignoring the correction of $\cO(\delta y_f^2)$ and higher.  
Put differently, the analysis is performed consistently at order $\cO(1/\Lambda^2)$, ignoring all  $\cO(1/\Lambda^4)$ effects (from $D$=8 operators, or from the square of $D$=6 contributions to the observables). 
This procedure leads to a Gaussian likelihood in the space of $\delta y_f$, in other words the corresponding $\chi^2$ function is a quadratic polynomial in $\delta y_f$. 
This polynomial can be unambiguously reconstructed given the central values, the 1~$\sigma$ uncertainties, and the correlation matrix  in \eref{3par}. 
Incidentally, the constraints change very little (by less than 20\%) if the EFT predictions are not expanded to a linear level, but instead the full non-linear dependence on $\delta y_f$ is retained. 
In such a case, the likelihood is highly non-Gaussian, but it can nevertheless be well approximated by a Gaussian likelihood in the parameter space region with $\delta y_f \ll 1$ which is preferred by the LHC Higgs data. 
The main qualitative consequence of using the full non-Gaussian likelihood is the existence of other nearly degenerate minima (in addition to the one described by \eref{3par}) where some $\delta y_f$ are large. 
For example,  the full likelihood contains minima with $\delta y_d \sim -2$ and/or $\delta y_e \sim -2$, as the Higgs data are currently very weakly sensitive to the sign of the bottom and tau Yukawa.  

The EFT likelihood defined by \eref{3par} can be recast into a  2HDM  likelihood by inserting the relation between $\delta y_f$ and the 2HDM parameters.
For example, in order to obtain constraints in the $c_{\beta-\alpha}$-$\tan \beta$ plane we need to read off from \eref{lhff} 
\beq
\delta y_f =  \sqrt{1-  c_{\beta-\alpha}^2} + \eta_f {c_{\beta-\alpha} \over \tan \beta} - 1 . 
\eeq 
Of course this procedure cannot be in any way better than deriving the limits  on $c_{\beta-\alpha}$ and $\tan \beta$ directly, 
as we did in \sref{2hdmfits}.
The purpose of this exercise is investigate how useful EFT is as a tool to constrain various BSM scenarios. 
The idea is that the LHC experiments present the EFT likelihood like the one in \eref{3par}, or a more general one depending on a larger number of EFT parameters that can be subsequently projected into the $\delta y_f$ subspace.
That likelihood function can be recast to quickly obtain constraints on a host of  BSM models.  
Our exercise is a case study for the validity of the EFT approach to LHC Higgs data, 
which allows one to understand limitations of the EFT and avoid possible pitfalls.

 \begin{figure}[tb]
\includegraphics[width=0.45 \textwidth]{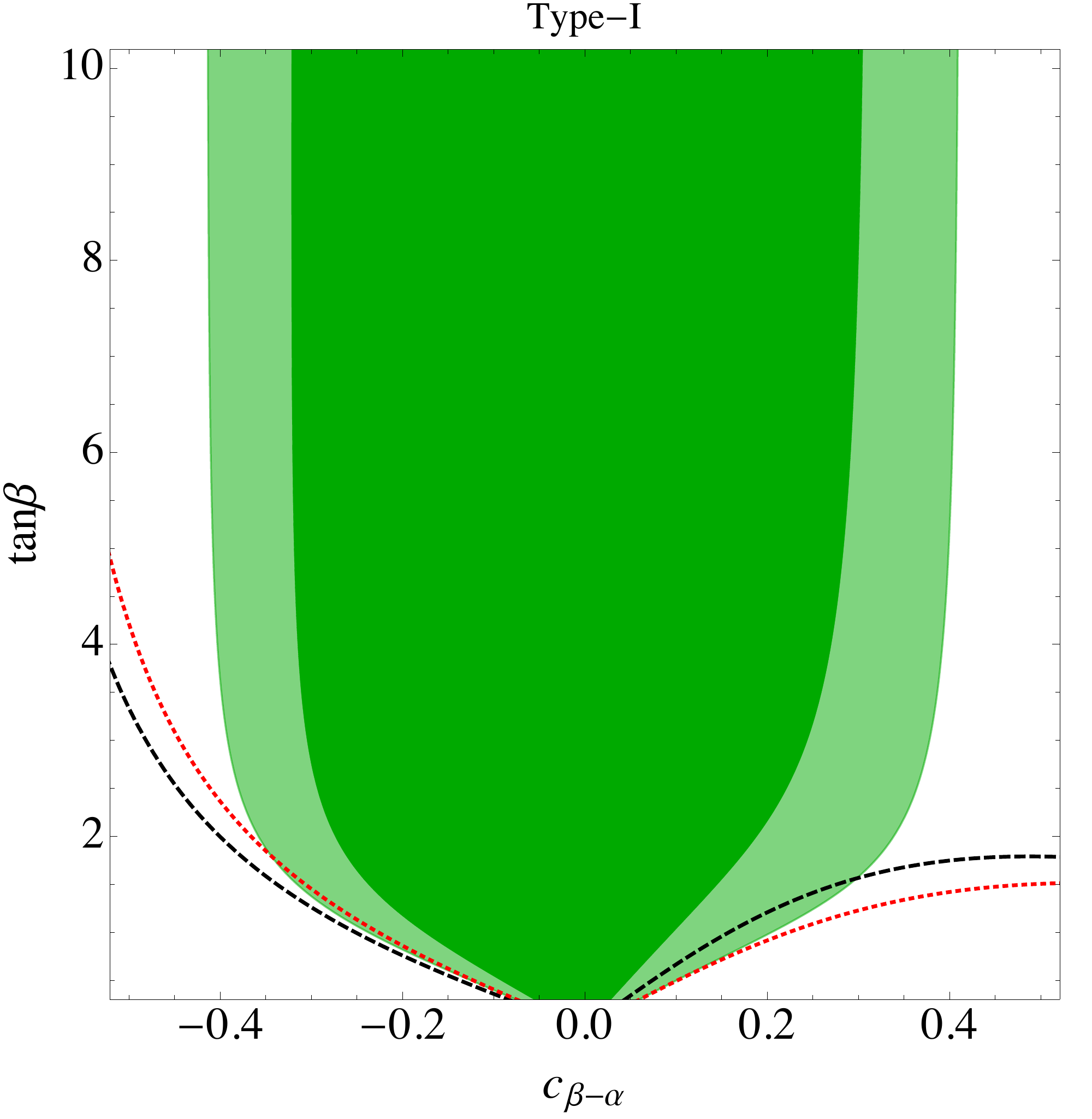}
\quad
\includegraphics[width=0.45 \textwidth]{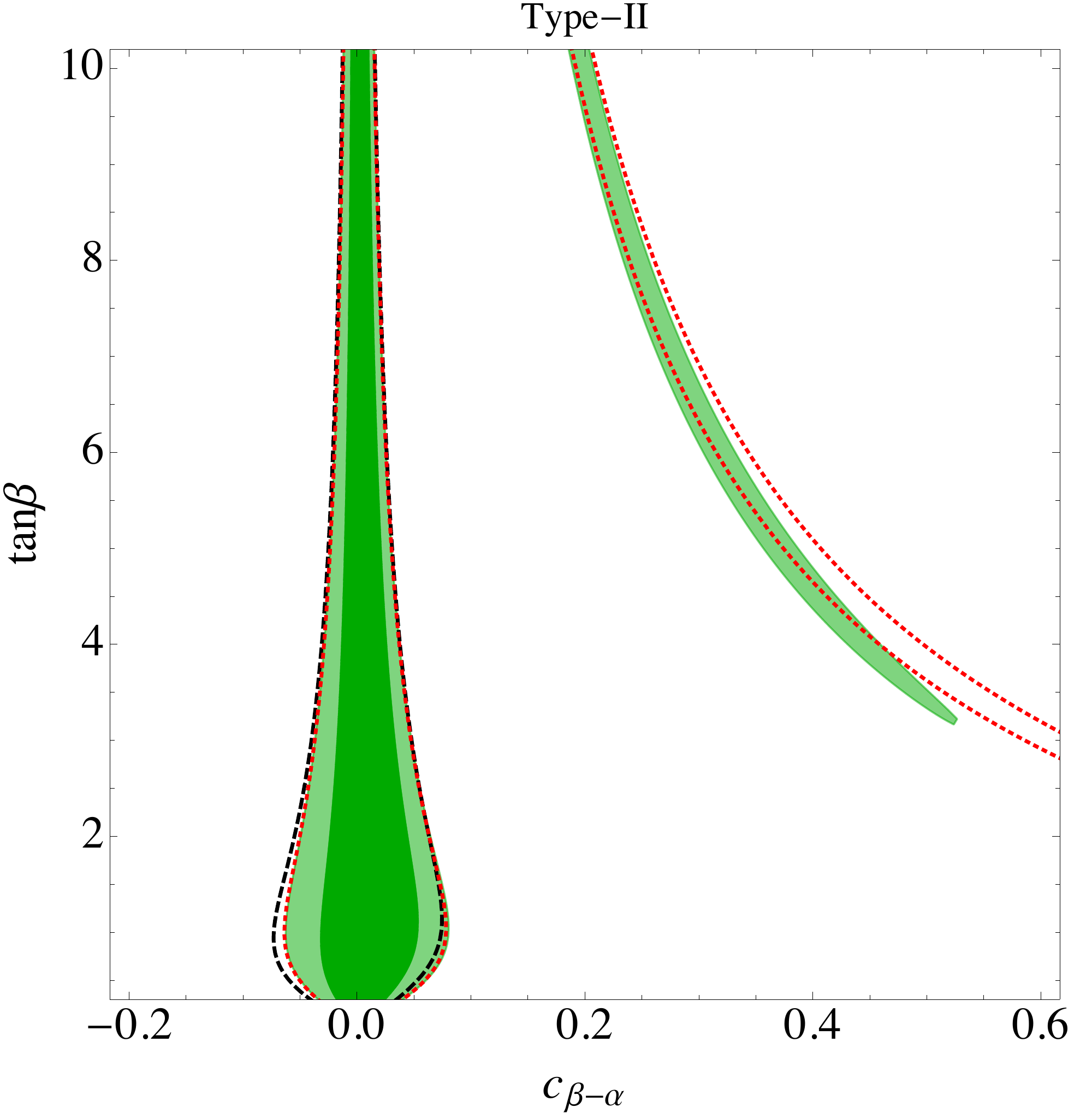}
\caption{
Constraints from LHC Higgs data on the parameter space of the type-I (left) and type-II (right) 2HDM.  
We show the 68\% (darker green) and 95\% (lighter green) CL region in the $c_{\beta-\alpha}$-$\tan \beta$ plane. 
We compare it with the 95\% CL region obtained by recasting the EFT limits in \eref{3par} which are derived from a Gaussian likelihood function (black dashed).  
We also show the boundary of the 95\% CL region obtained by recasting limits derived from  a non-Gaussian EFT likelihood where higher-order terms in $\delta y_f$ are kept (red dotted). 
}
\label{fig:EFT}
\end{figure}  
 
The results of our exercise are shown in \fref{EFT}. 
We compare the favored regions in the  $c_{\beta-\alpha}$-$\tan \beta$ plane for the type-I and type-II 2HDM  obtained in \sref{2hdmfits} with the ones deduced by using constraints on EFT parameters together with  the EFT-2HDM matching discussed earlier. 

Let's start the discussion with the type-II scenario. 
In this case, a recast of the Gaussian likelihood defined by \eref{3par} provides a very good description of the bulk of the favored region where  $c_{\beta - \alpha}$ is small. 
In that region, the LHC Higgs data force the deviations of the Yukawa couplings to be small, less than $\sim 30$\% of the SM value.
Such small deviations can be adequately  described by $D$=6 operators of the SM EFT, and the  $\cO(1/\Lambda^4)$ contributions to Higgs observables can be safely neglected. 
As the experimental precision increases, and assuming that no large deviations from the SM are reported, this conclusion will only be strengthened, and the agreement between the EFT and the complete description will further improve. 
On the other hand, we can see that the Gaussian EFT likelihood  completely misses the existence of the ``wrong-sign" Yukawa region. 
This is inevitable, as a Gaussian likelihood has only one minimum, and therefore it cannot capture other degenerate minima in the parameter space. 
The situation can be improved by complicating the description on  the EFT side, and instead including all higher-order terms in $\delta y_f$ in the likelihood function.  
Such a non-Gaussian likelihood is capable of describing multiple minima, including the  one in the ``wrong-sign" region where one or more $\delta y_f$ are smaller than  $-1$. 
Indeed, we can see in \fref{EFT} that using the non-Gaussian EFT likelihood qualitatively captures the shape of the ``wrong-sign" minimum, at least when $c_{\beta-\alpha}$ is not too large. 
Using the non-Gaussian likelihood also improves the agreement between the EFT and the direct 2HDM limits in the bulk region at small  $c_{\beta-\alpha}$.  

For type-X and type-Y the results are very similar as for type-II: the EFT description captures very  well the bulk of the favored parameter space with small $c_{\beta - \alpha}$, but it fails to capture the wrong-sign Yukawa region. 
Again, the latter problem can be addressed by using the non-Gaussian likelihood on the EFT side. 

In the type-I scenario a qualitatively new issue appears. 
In this case the EFT provides a good approximation  of the favored region for low $c_{\beta-\alpha}$ and $\tan \beta \lesssim 2$.
However, it completely misses the relevant physics at larger  $c_{\beta-\alpha}$ and  $\tan \beta$. 
Namely,  in the type-I  2HDM the LHC Higgs data imply an upper limit on $|c_{\beta -\alpha}|$, approximately $|c_{\beta -\alpha}| \lesssim 0.4$. 
At large $\tan \beta$, this limit is not driven by modifications of the Yukawa couplings, which are suppressed by $\tan \beta$, but rather by modifications of the Higgs couplings to $WW$ and $ZZ$. 
However, these appear only at $\cO(1/\Lambda^4)$ in the low-energy EFT of  the 2HDM, and are not included at all in the SM EFT truncated at $D$=6. 
In other words,  the type-I 2HDM at $\tan \beta \gg 1$ belongs to an  exceptional class of BSM scenarios that are not adequately described by a SM EFT with $D$=6 operators. 
Instead, in the  Higgs observables, the $D$=8 operators in the low-energy EFT (formally $\cO(\Lambda^{-4})$) may dominate over the $D=6$ ones (formally $\cO(\Lambda^{-2})$), as the latter  are suppressed by $\tan \beta$ and the former are not. 
This is an example of selection rules in the UV theory  modifying the naive power counting in the low-energy EFT. 
As  a consequence, the $D$=6 EFT approach in this case misrepresents the allowed parameter space of the type-I 2HDM at large $\tan \beta$. 
Note that the problem is not addressed by switching from a Gaussian to a non-Gaussian EFT likelihood. 
A more general low-energy approach is needed  to capture this scenario, for example the SM EFT truncated at the level of $D$=8 operators,  or a more phenomenological non-EFT approach.

\subsection{Parameter scans}

So far we have limited ourselves to studying the constraints on the 2HDM resulting from the LHC studies of the 125 GeV Higgs boson. 
These select  an interval(s) for the allowed values of the Higgs mixing angle $c_{\beta-\alpha}$, depending on the 2HDM scenario and on the value of the $\tan \beta$ parameter.
However, there exist further important constraints on the 2HDM.  
First of all, the neutral scalar and pseudo-scalar, and the charged partners of the Higgs boson are targeted by direct searches in high-energy colliders. 
Moreover,  the Higgs partners may contribute to electroweak precision observables, 
in particular to the $S$, $T$, and $U$ parameters \cite{Grimus:2008nb}, or to $Z \to b \bar b$ decays \cite{Haber:1999zh,Ferreira:2014naa}. 
Finally, the parameters of the Higgs potential should satisfy the theoretical constraints following from 
perturbative unitarity \cite{Kanemura:1993hm,Akeroyd:2000wc,Ginzburg:2003fe}, and the Higgs potential should be bounded from below \cite{Deshpande:1977rw}.
In the EFT approach, all of this information is not used. 
In particular, the heavy Higgses are integrated out from the spectrum.
The natural question then is whether the region in the  $c_{\beta-\alpha}$-$\tan \beta$ plane selected by the 125 GeV Higgs data can be realized in the full 2HDM given the existing constraints. 

\begin{figure}[tb]
\includegraphics[width=0.45 \textwidth]{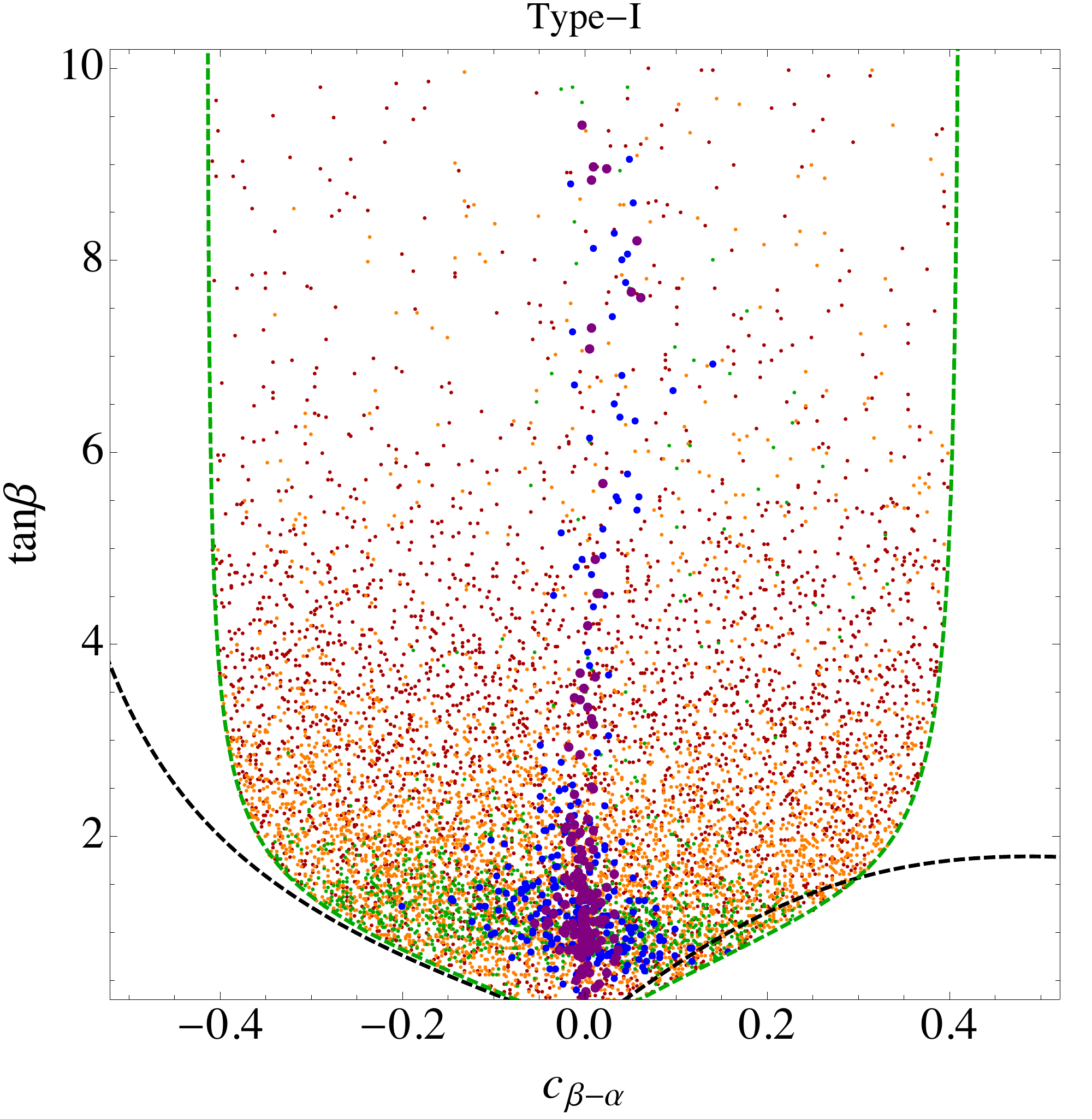}
\quad
\includegraphics[width=0.45 \textwidth]{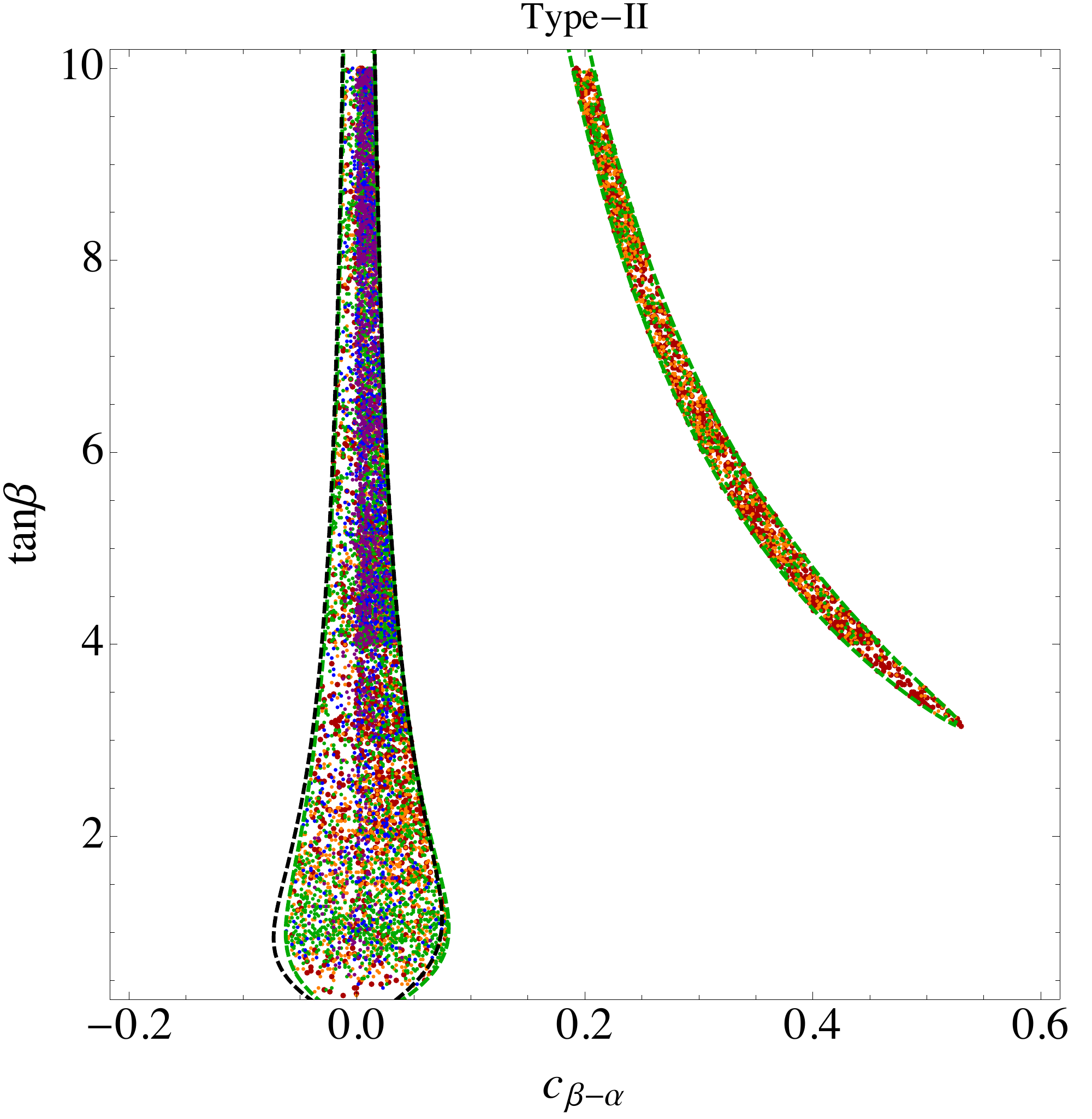}
\caption{
Scatter plot of the parameter space points of the 2HDM type-I (left) and type-II (right) scenario. 
The points satisfy constraints from perturbative unitarity and boundedness of the potential, electroweak precision observables,  and from the LHC analyses of the 125~GeV Higgs. 
The colors corresponds to different ranges for $m_A$: [125,200]~GeV~(red),  [200,400]~GeV~(orange),  [400,700]~GeV~(green),  [0.7,1]~TeV~(blue),  [1,2]~TeV~(purple).
For comparison, we also draw the contours of the 95~\% CL region favored by the Higgs data alone obtained using the direct approach (green dashed) or by recasting the Gaussian  EFT likelihood (black dashed). 
}
\label{fig:SCAN}
\end{figure} 

To address this question, we have performed scans of the 2HDM parameter space.
Our results are shown in \fref{SCAN}. 
We set $m_h = 125$~GeV, and generate points with $-\pi/2 \leq \alpha \leq \pi/2$, $0 \leq \tan \beta \leq 10$, $125~\gev \leq m_{H_0,A,H^+} \leq 2$~TeV.
For the type-II scenario we demand in addition $m_{H^+} \geq 480$~GeV, so as to satisfy the $b \to s \gamma$ constraints  \cite{Deschamps:2009rh,Hermann:2012fc,Eberhardt:2013uba,Ferreira:2014dya}.
In that range, we search for parameter points  that pass the 125~GeV Higgs constraints evaluated earlier, as well as the unitarity, boundedness,  and electroweak precision constraints. 
The latter constraints turn out to be non-trivial, in the sense that they eliminate a fraction of points that would pass the Higgs constraints alone. 
However, in our simulations they seem not to eliminate any particular value of $c_{\beta - \alpha}$ and $\tan \beta$ favored by the Higgs data. 
This is  known as  the {\em emmental effect}:  excluded regions in the multi-dimensional parameter space of the 2HDM do not show after a projection onto the two-dimensional $c_{\beta - \alpha}$-$\tan \beta$ plane.

For type-II, the bulk  of the allowed parameter space with $|c_{\beta - \alpha}| \ll 1$ contains points with extra scalar masses ranging from very heavy ($\gtrsim 1$~TeV) to very light ($\sim 125$~GeV),  corresponding to alignment with and without decoupling, respectively.  
That region is also recovered (to a good approximation)  by recasting the Gaussian EFT analysis of Higgs data into constraints on 2HDM parameters. 
Our scan shows that this entire region can be realized in the 2HDM  with all extra scalars decoupled at the LHC energies.
In such a case, the heavy states are not accessible directly, and their only observable effect is to modify the couplings of the 125~GeV Higgs. 
In the wrong-sign region, on the other hand, the extra scalars need to be relatively light, $m_{H_0,A,H^+}\lesssim 700$~GeV. 
This implies the heavy states are not decoupled at the LHC energies, and can be relevant for direct searches and resonant double Higgs production analyses.
Thus, while the wrong-sign region is perfectly consistent with  the current experimental data, the related LHC phenomenology is strictly speaking not amenable to an EFT description. 
We conclude that, for the type-II scenario, the SM EFT approach at $\cO(\Lambda^{-2})$ is adequate in the entire parameter space allowed by the experimental data and where the additional scalars are heavy. 
Similar conclusions hold for the type-X and type-Y models. 

The situation is somewhat different for the type-I scenario. 
As we discussed in \sref{eftfits}, the discrepancy between the full model and the EFT description is  important, especially at large $\tan \beta$.
The reason is that in this case  the numerically largest effects on the Higgs boson couplings are $\cO(\Lambda^{-4})$ and correspond to $D$=8 operators, whereas the formally leading  $\cO(\Lambda^{-2})$ effects, captured by the $D$=6 SM EFT, are suppressed by $\tan \beta$. 
This problem will always be present at large enough $\tan \beta$ even when  precision of Higgs measurements is improved significantly.
However, the scan in \fref{SCAN} shows that the parameter space where the two descriptions disagree about the Higgs couplings bounds is dominated by points with very light extra scalars. 
Thus, much like in the type-II case, most of the 2HDM parameter space where the EFT approach fails to deliver the correct bounds is anyway the one where the extra scalars do not decouple at the LHC energies.

\subsection{Discussion}

Working with the SM EFT one always needs to make a compromise between generality and simplicity.
In principle, the full EFT Lagrangian contains all information about the effects of heavy new physics on low-energy observables, but that information is encoded in an infinite number of parameters.    
The usual approach is to truncate the EFT expansion at the level of $D$=6 operators, which corresponds to retaining  the effects up to $\cO(\Lambda^{-2})$ in the new physics mass scale $\Lambda$.
If that is done consistently, that is the EFT predictions for the observables are expanded up 
$\cO(\Lambda^{-2})$ then low-energy measurements, such as the LHC Higgs signal strength observables, can be translated into a Gaussian likelihood for the $D$=6 EFT parameters. 
This allows for a very concise presentation of results, as a Gaussian likelihood is fully specified by the central values, 1~$\sigma$ uncertainties, and correlation matrix of the parameters. 
Thus, using the SM EFT at $\cO(\Lambda^{-2})$,  the large amount of data contained in multiple Higgs analyses at the LHC can be summarized by just a handful of numbers that can be later recast to provide  constraints on a large class of BSM scenarios. 

The question is how much  information about the UV physics is lost due to   these dramatic simplifications. 
This can be addressed quantitatively by comparing the performance of  complete UV models and the $D=6$ EFT  approximating the low-energy physics of those models. 
Our case study of 2HDM scenarios and their EFTs leads to a few interesting conclusions.   
First of all, the Gaussian likelihood provides a very good approximation of the new physics effects  in the bulk of the allowed parameter space. 
We however identified the exceptional situations where this is not the case:
\ben 
\item In the wrong-sign Yukawa regions of type-II, -X and -Y scenarios, where the relative corrections to the Yukawa bottom and/or tau couplings are large and cannot be properly described at  
$\cO(\Lambda^{-2})$. 
\item  For the type-I scenario at large $\tan \beta$, where the leading effects on the single Higgs production and decay come from $D$=8 operators in the EFT, which are by default neglected when  
the SM EFT is truncated at $D$=6.
\een  

One can always complicate the EFT framework such that it is capable of describing also these special cases.  
In particular, the wrong-sign region can be captured in the EFT if one works with the $D$=6 EFT but retains the higher order terms in $D$=6 parameters.  
Indeed, the Higgs signal strength observables depend also on the squares and higher powers of the $D$=6 EFT Wilson coefficients, which are formally  $\cO(\Lambda^{-4})$ or more suppressed.
These are crucial to properly describe the situation when new physics contributions to observables are comparable or exceed the SM ones. 
As we have shown, retaining these contributions allows one to approximately reproduce the wrong-sign regions in the 2HDM, at a price of introducing non-Gaussian terms into the likelihood. 
To cover the large $\tan \beta$ region of the type-I scenario the EFT Lagrangian would have to be extended to include $D$=8 terms.

Both of these complications would make it more challenging to perform EFT analyses   at the LHC and present their results.  
In our opinion, the Gaussian approach with the EFT Lagrangian truncated  at $D$=6 may be productive in the context of Higgs signal strength observables.  
This simple approach is sufficient in generic situations, while the special cases described above can be treated separately.  
Indeed, our parameter scans show that the special cases are always associated with the extra scalars being not much heavier than the 125 GeV Higgs boson, and therefore they should be probed directly using the complete model description and without passing through the EFT.

\section{Beyond 2HDM}
\setcounter{equation}{0} 
\label{sec:bthdm} 

The LHC measurements of the Higgs signal strength summarized in \tref{HIGGS_datarun12}  show some tension with the SM predictions.
On the one hand, there is an excess  in the $t \bar t h$ production mode appearing in several Higgs decay channels. 
On the other hand, the signal strength in the $h \to b \bar b$ decay channel is suppressed for several production modes. 
Assuming for a moment this is not merely a statistical fluctuation, the data may point to the Higgs-top (-bottom) coupling being $30\%$ larger (smaller) than in the SM. 
Within the 2HDM, it is straightforward to arrange the Higgs couplings to top quarks to be enhanced, and the Higgs coupling to bottom quarks to be simultaneously suppressed. 
This happens in the type-II and type-Y models at $\tan \beta \sim 1$ and $c_{\beta-\alpha} > 0$.
However, these regions of the parameter space are not favored by the global fits showed in \fref{THDM_1} or in \fref{THDM_2}.
More generally, in the 3-parameter EFT fit in \eref{3par} the SM point where all $\delta y_f = 0$
is not significantly disfavored,  with $\chi^2_{\rm SM} - \chi^2_{\rm min} \approx 2$. 
The reason is that increasing the Higgs-top coupling also increases the gluon fusion cross section via the 1-loop top triangle diagram contribution to the $g g \to h$ amplitude.
Since the measured total Higgs cross section (which is dominated by gluon fusion) agrees very well with the SM predictions, simply increasing the top-Higgs couplings is not preferred by global fits. 
{\em Decreasing} the Higgs-bottom coupling  is disfavored for similar reasons. 
As the Higgs width is dominated by decays to bottom quarks, a smaller  Higgs-bottom coupling increases the Higgs branching fractions (and thus the signal strength) into other final states. 
In a global fit, the gain from fitting better the suppressed $h \to bb$ channels is outweighed by overshooting the signal strength in the precisely measured $WW$, $ZZ$, and $\gamma \gamma$ final states.

The above discussion suggests a simple ad-hoc solution to improve the global fit in a theory with two Higgs doublets.
One can arrange additional contributions to the effective Higgs-gluon coupling beyond those from integrating the top quark and other SM fermions. 
If the sign of these contributions was opposite to that induced by the top, the new physics could cancel the effect of the increased Yukawa in the gluon fusion Higgs production cross section.
We can parametrize these new contributions  by adding a new term in the 2HDM Lagrangian 
\beq
\label{eq:cggdef}
\cL = \cL_{\rm 2HDM} +  c_{gg} {g_s^2 \over 4} {h \over v} G_{\mu \nu}^a  G^{\mu \nu,a},   
\eeq 
where $G_{\mu \nu}^a$ is the gluon field strength, and $g_s$ is the SM strong coupling. 
The parameter $c_{gg}$ encodes the effects  of heavy colored particles beyond the 2HDM  on the Higgs phenomenology.  
For example, integrating out a new color octet scalar $S^a$ of mass $m_S$ coupled to the Higgs sector via the interaction term $+ \lambda_S |H_1|^2 S^a S^a$,  one finds $c_{gg} = - {\lambda_S v^2 s_{\beta-\alpha}^2 \over 16 \pi^2 m_S^2}$.
Similar extensions of the 2HDM have been considered in the past, see e.g.  \cite{Manohar:2006ga,Degrassi:2010ne,Cheng:2016tlc}.

We first  employ the linearized EFT approach to see whether allowing the parameter $c_{gg}$ to vary freely can lead to an improvement of the Higgs fit. 
The Higgs boson couplings are those in \eref{PEL_lh} with non-zero $\delta y_f$ and $c_{gg}$ and the remaining coupling set to zero.  
With that assumption, the Run-1 and Run-2 Higgs data lead to the following constraints:
\beq
\label{eq:4par}
\bvec \delta y_ u \\ \delta y_ d \\\delta y_ e \\ c_{gg}  \evec  = 
\bvec 0.22  \pm 0.15  \\ -0.37  \pm 0.14  \\  -0.10 \pm 0.14 \\ 
-0.0042 \pm 0.0014 \evec, 
\quad \rho =  \left ( \ba{cccc} 
1 & 0.05 & -0.05 & -0.74   \\ 
0.05 & 1 & 0.40 & 0.56   \\ 
-0.05  &  0.40 & 1  &  0.30 \\ 
-0.74 & 0.56 & 0.30 & 1  \ea \right ).   
\eeq 
Now the preferred values of the EFT parameters are significantly away from the SM point.  
Indeed, we find $\chi^2_{\rm SM} - \chi^2_{\rm min} \approx 11$, which translates to the $2.3 \sigma$ preference for BSM.  
We  also checked that allowing for more free parameters in the EFT (e.g. $c_{\gamma \gamma}$) does not lead to further significant improvement of the fit.

\begin{figure}[t]
\includegraphics[width=0.45 \textwidth]{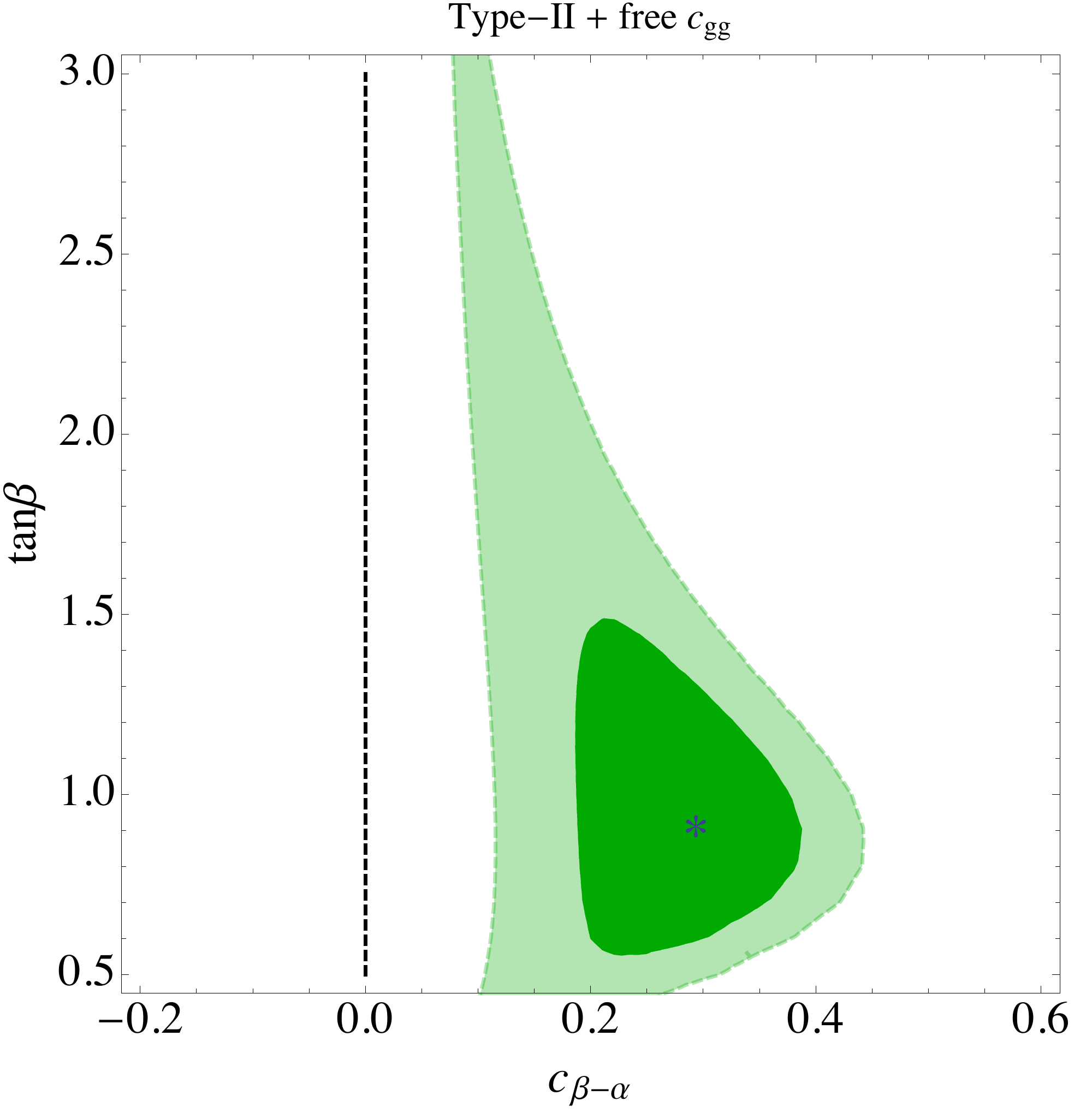}
\quad
\includegraphics[width=0.45 \textwidth]{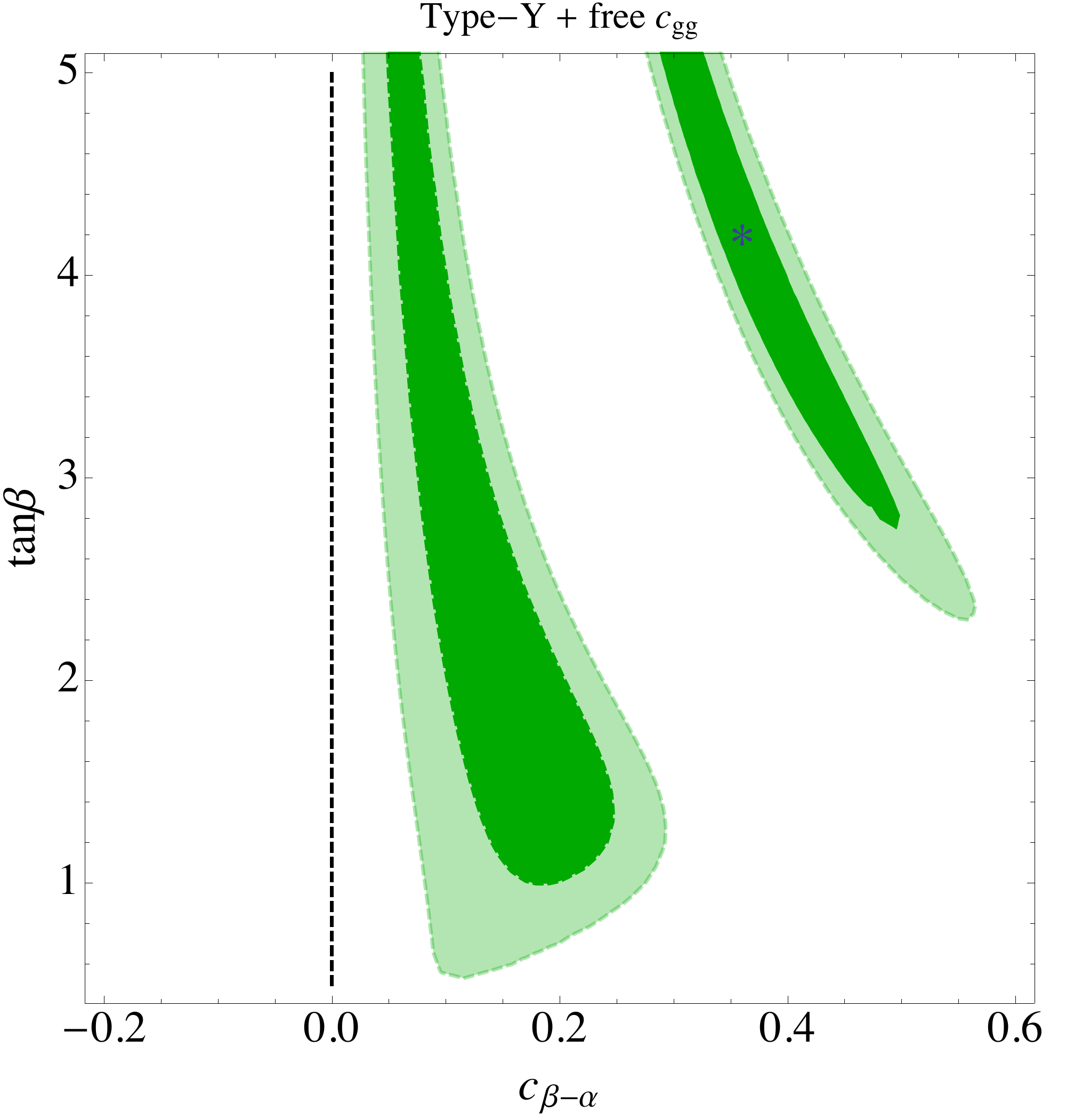}
\quad 
\caption{Constraints from LHC Higgs data on the parameter space of the type-II (left panel) and type-Y (right) 2HDM extended by the effective coupling in \eref{cggdef}.  
We show the 68\% (darker green) and 95\% (lighter green) CL region in the $c_{\beta-\alpha}$-$\tan \beta$ plane after marginalizing the likelihood  at each point  over the parameter  $c_{gg}$.
}
\label{fig:F2HDM}
\end{figure}

The EFT results  in \eref{4par} suggest that the Higgs fit can be improved also in the context of the type-II and type-Y scenario, once we allow for the new contributions to the Higgs-gluon coupling.  
This is indeed the case, as shown in \fref{F2HDM}.  
The best fit regions move away from the SM limit  where $c_{\beta-\alpha} = 0$ to  $c_{\beta -\alpha} > 0$ where $\delta y_u > 0$ and $\delta y_d <0$ are effectively generated.   

For the type-II case the best fit point occurs for $c_{\beta -\alpha} \approx 0.29$, $\tan \beta \approx 0.9$, and $c_{gg} \approx -4.0 \times 10^{-3}$,  and  has $\chi^2_{\rm SM} - \chi^2_{\rm min} \approx 15$. 
The minimum is slightly deeper than in the 4-parameter EFT fit because here we use the full (not Gaussian)  likelihood function.  
For such low $\tan \beta$ constraints from flavor physics become non-trivial and require $m_{H^+} \gtrsim 480$~GeV \cite{Deschamps:2009rh,Hermann:2012fc,Eberhardt:2013uba,Ferreira:2014dya}. 
Nevertheless, this limit does not pose consistency problems, as $c_{\beta -\alpha} \sim 0.3$ can be obtained with perturbative couplings in the scalar potential as long as $m_{H^+} \lesssim 1.5$~TeV. 
The preferred value of $c_{gg}$ requires a large contribution to the effective Higgs-gluon coupling from new particles, approximately one half (in magnitude) that of the top quark in the SM.
In the example with a scalar  octet  we need $\lambda_S \approx 11 \times (m_S/\tev)^2$, 
thus the octet needs to be below the TeV scale for  $\lambda_S$ to remain perturbative.  
Note that current LHC and Tevatron data still do not exclude fairly light colored particles in a model-independent way, see e.g. \cite{Cheng:2016tlc,Blum:2016szr} for a recent discussion. 
On the other hand, the approximate cancellation between all BSM contributions to the gluon fusion amplitude  does not have a natural explanation in this model, and should be considered an accident.

For the type-Y case the best fit point falls actually into the wrong sign region, at  $c_{\beta -\alpha} \approx 0.36$, $\tan \beta \approx 4.2$, $c_{gg} \approx - 2.6 \times 10^{-3}$ and has $\chi^2_{\rm SM} - \chi^2_{\rm min} \approx 14$.
However, it is not strongly preferred  ($\chi^2$ lower by just $0.3$) over  the local minimum 
 at  $c_{\beta -\alpha} \approx 0.17$, $\tan \beta \approx 1.8$, and $c_{gg} \approx -2.6 \times 10^{-3}$ where all Yukawas are positive. 
The higher $\tan \beta$ and lower $c_{\beta - \alpha}$ at the local minimum in the type-Y case are somewhat easier to accommodate than the best fit point for type-II. 
For example, in the scalar octet case we  need $\lambda_S \approx 7 \times (m_S/\tev)^2$,  and the flavor physics bounds on $m_{H^+}$ are not relevant for the preferred $\tan \beta$. 

For the case of type-I and type-X models we do not find any significant improvement of the fit after introducing the parameter $c_{gg}$. 
This is due to the fact that in these scenarios $\delta y_u = \delta y_d$, therefore one cannot simultaneously fit the enhanced $t \bar t h$ and suppressed $h \to b b$ signal. 

\section{Summary}

In this paper we discussed the validity of the SM EFT with $D$=6 operators as a low-energy theory for the 2HDM. 
Working consistently at $\cO(\Lambda^{-2})$ in the EFT expansion, the LHC Higgs signal strength measurements can be recasted into a Gaussian likelihood for the EFT Wilson coefficients. 
That likelihood can then be used to place constraints on the parameter space of various extensions of the SM, once the matching between the BSM model and its low-energy EFT is known.  
We applied this procedure for the case of the CP-conserving 2HDM, restricting to the tree-level matching.  
We then compared the resulting constraints on the $c_{\beta-\alpha}$-$\tan \beta$ plane with those derived directly without passing through the EFT.  
We find that, in the bulk of the allowed parameter space of the 2HDM where $c_{\beta-\alpha}$ is small,  the Gaussian likelihood approximates very well the effects  of the new scalars on the Higgs phenomenology. 
In those regions, the SM EFT truncated at $D$=6 provides a valid description of the 2HDM phenomenology, as long as the extra scalars are heavy enough such that they do not appear on-shell in LHC Higgs observables.  

However, we also identified the situations where our EFT procedure miscalculates the impact of the 2HDM  on Higgs physics, even when $\Lambda \gg m_h$.  
One occurs when some SM Yukawa coupling receives corrections that are comparable to its SM value, which happens in particular in the wrong-sign Yukawa regions. 
Another occurs for the type-I scenario at large $\tan \beta$, where the leading 2HDM effects on Higgs phenomenology are encoded in $D$=8 operators of the low-energy EFT.   
These two exceptions are important to keep in mind when EFT results are interpreted as constraints on BSM, as they are representative of a wider class of models. 
It is possible to generalize the EFT approach such that it becomes adequate also in the above situations,  but that would come at the price of a greater complexity of the analysis and a less transparent presentation. 

We also applied the EFT approach to investigate what deformations of the SM Higgs couplings are needed to improve the fit to the Higgs data.  
According to \eref{4par}, this requires simultaneously  1) increasing the top Yukawa coupling, 2) decreasing the bottom Yukawa coupling, and 3) inducing the contact interaction of the Higgs boson with gluons. 
We discussed how these modifications can be realized in the 2HDM extended by new colored particles coupled to the Higgs.
Future analyses of the LHC data from the 2016 run will tell whether the current small tension between the measurements and the SM predictions is just due to a statistical  fluctuation, or due to new physics contributions to the Higgs boson couplings.  

\section*{Acknowledgments}

This work was supported by the {\em Partenariats Hubert Curien} programme PESSOA  under project  33733UH, and by the Portuguese FCT under project 441.00 of Portugal/France cooperation program PESSOA 2015/2016. 
A.F is partially supported by the ERC Advanced Grant Higgs@LHC and by the European UnionÕs Horizon 2020 research and innovation programme under the Marie Sklodowska-Curie grant agreements No 690575 and  No 674896.
The work of J.C.R. and J.P.S. is supported in part by the Portuguese \textit{Funda\c{c}\~{a}o para a Ci\^{e}ncia e Tecnologia} (FCT) under contract UID/FIS/00777/2013.

\appendix
\renewcommand{\theequation}{A.\arabic{equation}} 
\setcounter{equation}{0}


\bibliographystyle{JHEP} 
\bibliography{twohiggspaper}

\end{document}